\documentclass[twocolumn,prc,aps,longbibliography]{revtex4-1}
\usepackage{amssymb}
\usepackage{graphicx}
\usepackage{dcolumn}
\usepackage{amsmath}
\usepackage{multirow}
\usepackage{booktabs}

\begin{document}
\title{Invariant-mass spectroscopy of $^{18}$Ne, $^{16}$O, and $^{10}$C excited states formed in neutron transfer reactions.}
\author{R. J. Charity}
\author{K. W. Brown}
\author{J. Elson}
\author{W. Reviol}
\author{ L. G. Sobotka}
\affiliation{Departments of Chemistry and Physics, Washington University, 
St.~Louis, Missouri 63130, USA.}
\author{W. W. Buhro}
\author{Z. Chajecki}
\thanks{Pressent Address: Department of Physics, Western Michigan University, Kalamazoo, Michigan, 49008, USA.}
\author{W. G. Lynch}
\author{J. Manfredi}
\author{R. Shane}
\author{R. H. Showalter}
\author{M. B. Tsang}
\author{D. Weisshaar}
\author{J. Winkelbauer}

\affiliation{National Superconducting Cyclotron Laboratory and 
Department of Physics and Astronomy, Michigan State University, 
East Lansing, MI 48824, USA.}

\author{S. Bedoor}
\affiliation{Department of Physics, Western Michigan University, Kalamazoo, 
Michigan 49008, USA.}
\author{D. G. McNeel}
\author{A. H. Wuosmaa}
\affiliation{Department of Physics, Western Michigan University, Kalamazoo, 
Michigan 49008, USA. and Department of Physics, University of Connecticut, Storrs, Connecticut 06269, USA.}

\date{\today}

\begin{abstract}
Neutron transfer reactions with fast secondary beams of $^{17}$Ne, $^{15}$O, and $^9$C have been studied with the HiRA and CAESAR arrays. Excited states of $^{18}$Ne, $^{16}$O, and $^{10}$C in the continuum have been identified using invariant-mass spectroscopy. The best experimental resolution of these states is achieved by selecting events where the decay fragments are emitted transverse to the beam direction. We have confirmed a number of  spin assignments made in previous works for the negative-parity states of $^{18}$Ne. In addition we have found new higher-lying excited states in $^{16}$O and $^{18}$Ne, some of which fission into two ground-state $^8$Be fragments. Finally for $^{10}$C, a new excited state was observed. These transfer reactions were found to leave the remnant of the $^9$Be target nuclei at very high excitation energies and maybe associated with the pickup of a deeply-bound $^9$Be neutron.  
 
\end{abstract}
\maketitle

\section{Introduction}
Invariant-mass spectroscopy with fast radioactive beams has proven a valuable tool for studying the structure of light exotic isotopes near the 
drip lines. With the High Resolution Array (HiRA) \cite{Wallace:2007}, we have focused our studies on states produced in nucleon knockout reactions for isotopes near and beyond the proton drip line \cite{Charity:2010,Charity:2011,Brown:2014,Brown:2014a,Brown:2015}. However in the same experiments, we also obtained data for a number of other reactions types \cite{Charity:2011,Brown:2017}. In this work we will report on levels obtained from neutron-transfer reactions with fast $^{17}$Ne, $^{15}$O, and $^{9}$C secondary beams using experimental data sets for which knockout results have  already published. One advantage of the invariant-mass technique is its selectivity to the decay channel. This allows one to isolate small cross sections associated with exotic exit channels and determine branching ratios  in decays.

The experimental technique will be validated by studying the well-known spectroscopy of $^{16}$O states which can be produced  with the $^{15}$O beam. In particular we will look at the $\alpha$-particle  branching ratio for the $J^{\pi}$=2$^{+}_{3}$ level which is important to determine its isospin mixing with the neighboring $J^{\pi}$=2$^{+}_{2}$ level \cite{Leavitt:1983}.
 With the $^{17}$Ne beam, we will look at the low-lying levels of $^{18}$Ne. 
The structure of $^{18}$Ne has attracted considerable interest due to its importance  for the resonant component of the 
$^{14}$O($\alpha$,$p$)$^{17}$F and $^{17}$F($p$,$\gamma$)$^{18}$Ne reactions in astrophysics \cite{Hahn:1996,Chipps:2009,Hu:2014}. In the course of such studies, Hahn \textit{et al.} \cite{Hahn:1996}  produced an evaluated  level scheme for this isotope and made spin assignments based on the level widths, cross sections and angular distributions in various reactions, and Thomas-Ehrman shifts relative to the mirror $^{18}$O system. Due to the selectivity of transfer reactions, only  levels of certain spins and parity will be strongly populated with a $^{17}$Ne beam and this can be used to check the spin assignments of Hahn \textit{et al.}   In addition for all three projectiles, we will look for previously unobserved  higher-lying excited states. Here the power of the invariant-mass technique will allow us to observe highly-fragmented decay channels with interesting decay modes. 

Our main interest is the low-lying particle-unstable states formed by neutron capture to the $p$ and $sd$ shells. However from semi-classical models of this process \cite{Brink:1972}, transfer of a nucleon to such orbitals  with fast beams ($E/A$=60-70~MeV) is poorly matched in terms of linear and angular-momentum transfer leading to small cross sections. Moreover, transfer reactions also have 
selectivity to structures with single-particle-like configurations and can be used to probe such structures and constrain models. Indeed at 
lower energies where linear and angular momentum are better matched, transfer reactions such as ($d$,$p$) have contributed significantly to this area using the missing-mass technique. Indeed, such cases are 
amenable to simple reaction theory (Distorted Wave Born Approximation for instance) and spectroscopic strengths and spin assignments can be inferred from the detected cross sections and angular distributions. However with fast secondary beams, the missing-mass technique requires thinner targets than those typically used with the invariant-mass technique. In addition, because of the large phase space of these secondary beams, beam tracking is required for the determination of absolute angles, whereas relative angles are only important in the invariant-mass 
technique, which in HiRA, are almost insensitive to the size of this phase space.  In this work we will explore the role that the invariant-mass technique can play in these transfer reactions and present its advantages and disadvantages. Finally this work is complementary to recent studies using $\gamma$-ray spectroscopy following transfer reactions with fast secondary beams where the final projectile-like fragment is detected in a spectrometer \cite{Gade:2007,Gade:2011,Gade:2016}.

\section{EXPERIMENTAL METHOD}
\label{sec:method}
The data presented in this work was obtained from experiments performed at the Coupled Cyclotron
 Facility at the National Superconducting Cyclotron Laboratory at Michigan 
State University. Details of these experiments have been described in 
Refs.~\cite{Brown:2014,Brown:2014a,Brown:2015} and only a brief description will be given here. A secondary beam of intensity 1.5$\times$10$^5$ pps was 
obtained from the fragmentation of an $E/A$=170-MeV $^{20}$Ne primary beam 
(80 pnA). This beam contained $^{17}$Ne (11\%)  and $^{15}$O (80\%) with energies 
in the center of a 1-mm-thick Be target of $E/A=$58.2 and 48.1 MeV, respectively.
In a separate experiment, a secondary beam of intensity 9$\times$10$^4$~pps was obtained from an $E/A=$ 150-MeV $^{16}$O primary beam (175 pnA). This beam contained  $^{9}$C at the 52\% level with an energy in the center of the same target of $E/A$=64.6 MeV. The other main component of this beam was  $^{6}$Li.

Charged particles produced from reactions with the target were detected in the High Resolution Array (HiRA) \cite{Wallace:2007} consisting of 14 $\Delta E-E$ telescopes arranged around the
beam to cover zenith angles from $2^{\circ}$ to $13.9^{\circ}$. The double-sided
Si strip $\Delta E$ detectors permitted accurate determination of the scattering
angles of the detected fragments. The heavier fragments ($A>$10) were only identified in the central two telescopes where the $\Delta E$ strips were set up with dual gains. 
Energy calibrations of the CsI(Tl) $E$ detectors  were achieved using a series of cocktail beams
including  $E/A$=55 and 75~MeV protons and $N$=$Z$ fragments, and $E/A$=73.4 and 95.2~MeV $^7$Be fragments.  Other fragments such as $^{15}$N and $^{17}$F have only a single calibration point each at $E/A$=40.1 and 51.3 MeV, respectively. In these cases, we use the calibration point to define  effective thicknesses  of the Si $\Delta E$ detectors and then use energy-loss tables  \cite{Ziegler:1985} to determine $E$ from the $\Delta E$ measurement.  
The relative locations of each HiRA telescope and the
target were determined very accurately using a Coordinate Measurement Machine
 arm.

The CAESAR (CAESium iodide ARray) detector  \cite{Weisshaar:2010} was positioned to  surround the target in order to detect  $\gamma$ rays emitted in coincidence with charged particles. For this experiment, the array consisted of 158 CsI(Na) crystals covering polar angles between 57.5$^\circ$ and 122.4$^\circ$ in the laboratory frame with complete azimuthal coverage. The first and last rings of the full CAESAR array were removed due to space constraints.

For the normalization of cross sections, the number of beam particles was determined by counting using a thin plastic-scintillator foil placed in the focal point of the A1900 fragment separator. For the $^{17}$Ne-$^{15}$O beam, the loss in the beam flux due to its transport to the target and the relative contribution from each beam species was determined by temporarily placing a CsI(Tl) detector just after the target position. These fluxes were also corrected for the detector dead time measured with a random pulse generator. No similar calibrations was performed for the $^9$C beam. Here we rely on a previous experiment with the same beam energy, target, and detector setup where a similar calibration was performed \cite{Charity:2011}. Normalization of cross sections in the present case was determined by reproducing the value for $^8$C$_{g.s.}$ from the previous experiment.  The uncertainties quoted for the cross sections in remainder of this work are statistical only. In addition to these, there is also a systematic uncertainty of $\pm$15\% for the  $^{18}$Ne and $^{16}$O  states and  $\pm$20\%  for the $^{10}$C states.  
 
\section{Invariant-mass Method}

For a  group of detected fragments believed to be the decay products of a nuclear level, we can calculate its excitation energy as
\begin{equation}
E^{*}_{n\gamma} = E_{inv} - E_{g.s}
\end{equation} 
where $E_{inv}$ is the invariant mass of the fragments and $E_{g.s.}$ is the ground-state mass of the decaying nucleus. However, the quantity $E^*_{n\gamma}$ is only the true excitation energy if no $\gamma$-rays were emitted in the decay. For example, the particle decay of a state may leave one or both of the decay fragments in particle-bound excited states which subsequently $\gamma$ decay. In such cases, the true excitation energy is obtained by adding the $\gamma$-ray energies, i.e.,  
\begin{equation}
 E^* = E^*_{n\gamma} + \sum_i E^{\gamma}_i .
\end{equation}
The use of the CAESAR $\gamma$-ray array allows us to identify such cases and apply this correction.

\begin{figure}
  \includegraphics*[scale=.4]{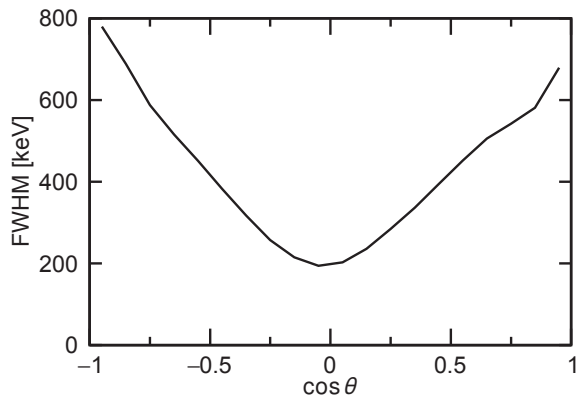}
  \caption{ Simulated $^{18}$Ne$\rightarrow p+^{17}$F resolution expressed as the FWHM of the 
invariant-mass distribution for a level with zero intrinsic width and $E^*$=5.135 MeV. Results are shown as a function of the $\theta$, the emission angle of the proton in the $^{18}$Ne$^*$ frame where $\theta$=0$^{\circ}$ corresponds to emission along the beam axis.}

\label{fig:FWHM}
\end{figure}

The experimental apparatus is only sensitive to particle decays of projectile-like states which are produced at laboratories angles close to the beam axis ($\theta_{lab} <$ 10$^\circ$).  For two-body decays where the invariant mass can be determined solely from the relative velocity between the two fragments, the experimental resolution depends very strongly on the decay direction. For example, Fig.~\ref{fig:FWHM} shows the simulated resolution  (App.~\ref{sec:MC}) expressed as a FWHM of the invariant-mass peak for the decay $^{18}$Ne$\rightarrow p$+$^{17}$F with an excitation energy of 5.135~MeV and zero intrinsic width.
The angle $\theta$ is the emission angle of the proton in the $^{18}$Ne$^*$ center-of-mass frame with $\theta$=0$^\circ$ corresponding to emission along the beam axis. 
This strong angular dependence reflects the fact that we have excellent relative-angle resolution, but poorer energy resolution, and the relative contribution of these to the total resolution is strongly $\theta$-dependent. In both cases, these resolutions are dominated by the effect of the thick target.
For the relative-angular resolution, it is the  small-angle scattering of the decay products in the target material which is important, while for the energy resolution, the uncertainty in the interaction depth in the target leads to an uncertainty in the energy loss of the decay fragments  as they leave the target.

\begin{figure}
  \includegraphics*[scale=.4]{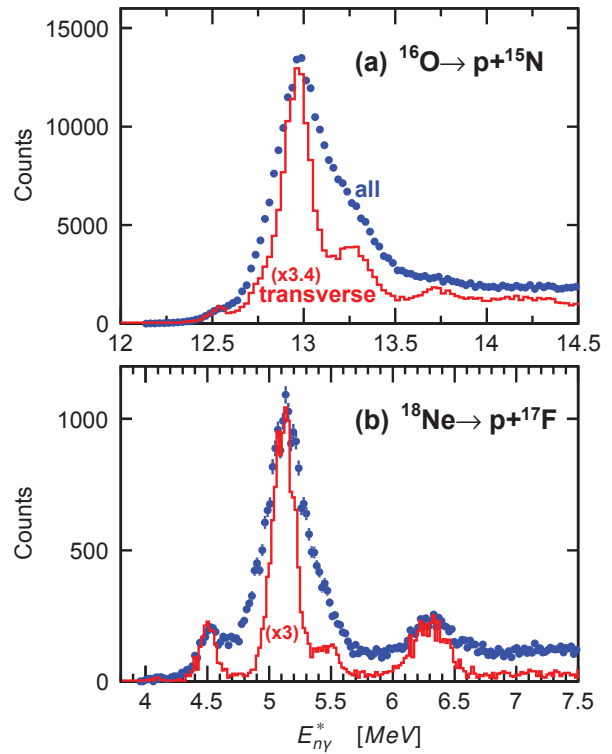}
  \caption{ Experimental excitation-energy spectra obtained with 
the invariant-mass method from detected (a) $p$+$^{15}$N and (b) $p$+$^{17}$F
events. The data points were obtained from all detected events, while the 
histrograms are for transverse decays only ($\left|\cos\theta\right|<0.2$).} 
\label{fig:trans}
\end{figure}

For transverse decays ($\cos\theta\sim$0),  uncertainties in the 
energies of the detected fragments act perpendicular to the decay axis and thus only contribute to the invariant-mass uncertainty in second order. In this case, the experimental resolution is dominated by the angular resolution. On the other hand for longitudinal decays ($\left|\cos\theta\right|\sim$1), the angular uncertainty contributes in second order and the experimental resolution is now dominated by the contribution from the energy. If there are enough statistics, it is clearly advantageous to restrict the analysis to events which decay transversely.
For example, Fig.~\ref{fig:trans} shows the inclusive (data points) and transverse-gated ($\left|\cos\theta\right|<$0.2, histograms) invariant-mass spectra for detected $p$+$^{15}$N and $p$+$^{17}$F events.  Both spectra show a number of peaks associated with $^{16}$O and $^{18}$Ne levels and our ability to resolve and identity these is clearly superior with the transverse gate. The transverse  gate $\left|\cos\theta\right|<$0.2 will be used in the following work unless otherwise specified.

 For similar reasons, the transverse-gated spectra also have reduced sensitivity to errors in the CsI(Tl) energy calibrations, thus reducing the systematic uncertainty 
in the  fitted peak energies.  To estimate the magnitude of this uncertainty we have fitted nine invariant-mass peaks associated with proton decay of $^{12,13,14,15}$N and $^{14,15}$O levels which have small intrinsic widths and their decay energies are well known. The weighted mean deviation from the ENSDF \cite{ENSDF} decay energies is -1.5(33)~keV. Thus we chosen a 2$\sigma$ deviation of 6.6~keV as a reasonable choice for this systematic uncertainty.

\section{$^{16}$O EXCITED STATES}
Neutron pickup by the $^{15}$O beam provides an excellent test of our understanding of transfer reactions at these higher energies as the $^{16}$O states of interest are well characterized and one can compared to lower-energy data from the mirror reaction, proton transfer to $^{15}$N \cite{Bohne:1971,Bohne:1972}.  
The ground-state configuration of $^{15}$O consists predominantly of a neutron hole in the $p$ shell. In neutron-transfer reactions, the lower-energy states are  produced by either filling this hole and making a $J^{\pi}$=0$^+$ state, or, by capturing the  neutron into the $sd$ shell. Of these possibilities, neutron capture to either the $d_{5/2}$ or $d_{3/2}$ level forming $J^{\pi}$=1$^-$, 2$^-$, or 3$^-$ states will have the smaller momentum mismatch and thus are expected to produce the largest cross sections at these energies. Capture to the $pf$ shell will generally produce states of larger excitation energy where the level density  increases and our experimental 
resolution is poorer  making it generally more difficult to isolate and identify them.

\begin{figure}
  \includegraphics*[scale=.43]{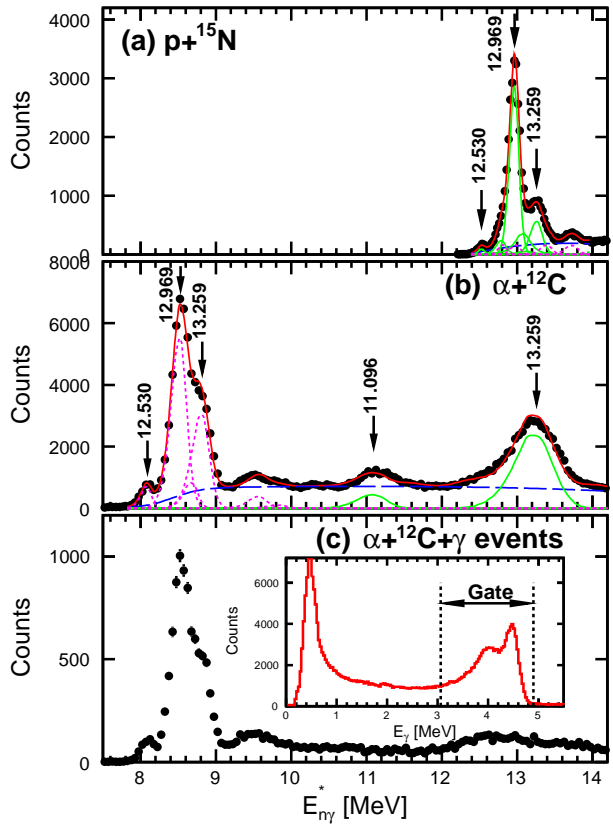}
  \caption{Distribution of $^{16}$O excitation energy reduced by the total energies of emitted  $\gamma$ rays  for events detected with the $^{15}$O beam. (a)+(b) Data points show the experimental distribution for all detected $p$+$^{15}$N and $\alpha$+$^{12}$C pairs, respectively. The solid-red curves show  fits to these distributions using known $^{16}$O levels. The individual contributions from these  levels are shown as the solid curves for decays to the respective ground states. In (b), decays to the first excited state of $^{12}$C are indicated by the dotted curves.  Background contributions (dash-blue curve) was also included in the fit.  (c) Excitation-energy spectrum gated on $\gamma$-rays from the decay of the first excited state of $^{12}$C. The inset shows the Doppler-corrected $\gamma$ spectrum measured in coincidence with $\alpha$+$^{12}$C pairs and the gate used to select $\gamma$-rays from the decay of this excited state.}

\label{fig:a12C}
\end{figure}

Invariant-mass spectra  for the $p$+$^{15}$N and $\alpha$+$^{12}$C transverse decay channels of $^{16}$O formed with the $^{15}$O beam are plotted in Figs.~\ref{fig:a12C}(a) and \ref{fig:a12C}(b). The observed peaks for $p$+$^{15}$N are all associated with decay to the ground-state of $^{15}$N apart from the highest-energy one ($E^*_{n\gamma}\sim$13.7~MeV) which will be discussed later (Sec.~\ref{sec:higher}).
 The $\gamma$-ray spectrum in coincidence with the detected $\alpha$+$^{12}$C pairs 
is shown in the inset in Fig.~\ref{fig:a12C}(c) where a peak associated with 
the  $E_{\gamma}$=4.438-MeV $\gamma$ ray from the decay of the first-excited state of $^{12}$C is visible.  Below this, the first escape peak is also clearly evident. Using the $\gamma$-ray gate indicated 
in the inset of Fig.~\ref{fig:a12C}(c) which encompasses both peaks, the resulting $^{16}$O excitation-energy spectrum is shown Fig.~\ref{fig:a12C}(c). 
 Comparing  this $\gamma$-gated and  the inclusive spectra  of Fig.~\ref{fig:a12C}(b), one finds both are almost identical in shape below $E^*$=10~MeV but not above and thus the lower-energy peak structures  must be associated with decays to the first excited state of $^{12}$C, while the higher-energy structures 
observed in Fig.~\ref{fig:a12C}(b) are associated with decays to the ground state.

Both the $p$+$^{15}$N and $\alpha$+$^{12}$C invariant-mass spectra have been fitted with peaks from the $^{16}$O levels that were observed in the lower-energy proton-transfer experiments with $^{15}$N targets \cite{Bohne:1971,Bohne:1972}. The peak energies and intrinsic widths were fixed to their values in  \cite{ENSDF}, while their intensities and a smooth background are varied to reproduce the data. Detector resolution is included via the Monte Carlo simulations (App.~\ref{sec:MC}).
The results
 are shown by the solid curves (red) with individual components indicated by the solid (green) curves for decay to the ground state or dashed (magenta) curves for decay to the excited state.  Note that for the $\alpha$+$^{12}$C$_{g.s.}$ decay channel, no  $J^{\pi}$=0$^-$, 2$^-$ levels are considered  
as such decays would violate parity conservation. These fits show that both spectra are dominated by the decay of  two $T$=1 states: the $J^{\pi}$=3$^-_{3}$ state at $E^*$=13.259~MeV observed in the $p$+$^{15}$N, $\alpha$+$^{12}$C$_{g.s}$, and $\alpha$+$^{12}$C$_{4.438 \textrm{MeV}}$ exit channels and a $J^{\pi}$=2$^-_{3}$ state ($E^*$=12.969~MeV) observed in the $p$+$^{15}$N and $\alpha$+$^{12}$C$_{4.438 \textrm{MeV}}$ channels. In addition the $T$=0, $J^{\pi}$=2$^-_{2}$ state at $E^*$=12.530~MeV is observed at  lower intensity in the $p$+$^{15}$N and $\alpha$+$^{12}$C$_{4.438 \textrm{MeV}}$ channels. Finally there is evidence for a peak at $E^*$=11.096~MeV in the $\alpha$+$^{12}$C$_{g.s.}$ channel at low yield which corresponds to a $J^{\pi}$=3$^+$ state, involving the capture of a $f$-shell neutron. 
The fits confirm our expectation that states formed by neutron capture to the $d$-orbital will dominate. Also, the experimental spectra were fit without any significant contribution from the $E^*$=10.957 and 12.796~MeV $J^{\pi}$=0$^-$ states and the $E^*$=12.440 and 13.090~MeV, $J^{\pi}$=1$^-$ states which all involve capture to the second 
$s_{1/2}$ level even though their spectroscopic factors are significant \cite{Bohne:1972}. This suppression of $s_{1/2}$ capture is consistent with a larger momentum mismatch at these higher bombarding energies.

\subsection{Branching Ratio of $J^{\pi}$=2$^{-}_{3}$  Level }
The $J^{\pi}$=2$^-$ states at $E^*$ = 12.530~MeV ($T$=0) and 12.969~MeV ($T$=1) are close enough in energy that there is some isospin  mixing. The magnitude of this mixing can be determined fom their $\alpha$-particle reduced widths \cite{Leavitt:1983}. However there is a disagreement in the value of the $\alpha $ partial width or branching ratio for the ($T$=1) 12.969~MeV state. Historically, the first information on this branching ratio is from the compilation of  Ajzenberg-Selove \cite{Selove:1977} giving $\Gamma_{\alpha1}/\Gamma$ = 0.36(5). 
This value was referenced to a paper of Rolf and Rodney \cite{Rolf:1974} where the branching ratio is not given or discussed, so details of the derivation of this value are unknown. Later  Leavitt \textit{et al.} measured a similar value of  0.37(6) from which they extracted a mixing parameter and the charge-dependent matrix element \cite{Leavitt:1983}. Subsequently Zijderhand and Van der Leun \cite{Zijderhand:1986} measured a smaller value of 0.22(4) which is in disagreement with the two previous measurements. It is this final value that is listed in the current ENSDF evaluation \cite{ENSDF}.

We have extracted the relative strength of the proton and alpha branches for transverse decay only. For longitudinal decay, the experimental resolution is much poorer making it very difficult to separate the 12.969 and 13.259~MeV states in both exit channels. If we assume the decay angular distributions are isotropic, then 
$\Gamma_{\alpha1}/\Gamma$=0.49 which is  larger than all of the other measurements.

However to the extent that these transfer reactions are peripheral, then the orbit of the neutron before transfer in the target and after transfer in the projectile should should lie predominantly in the reactions plane. As such the spin vector of the $^{16}$O excited states may show an overall alignment perpendicular to the beam axis.
 A minimum value of the branching ratio can be obtained using angular distributions calculated assuming the $J^{\pi}$=2$^-$ state has maximal alignment, i.e. $M$=0 with the beam axis as the quantization axis. Taking the proton decay as a $d_{5/2}$ emission \cite{Bohne:1972}, we obtain  $\Gamma_{\alpha1}/\Gamma>$0.32 which is inconsistent with  Zijderhand and Van der Leun, but consistent with the other measurements.

\subsection{$p$+$^{15}$N + $\gamma$  Exit Channels}
\label{sec:higher}
\begin{figure}
  \includegraphics*[scale=.43]{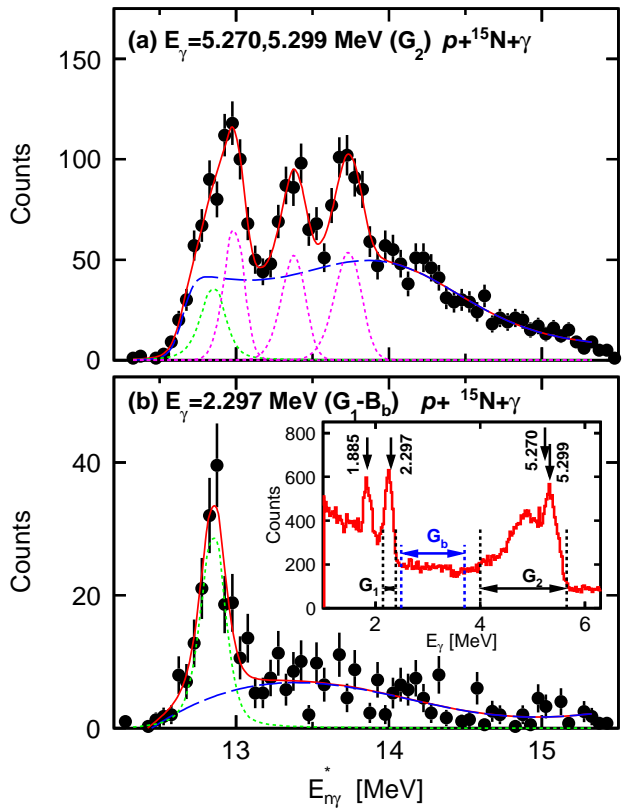}
  \caption{ Distribution of $^{16}$O excitation energy reduced by the total energies of $^{15}$N $\gamma$ rays  for  $p$+$^{15}$N pairs detected with the $^{15}$O beam. Data points show the experimental distributions, while solid-red curves show fits to these data. The dash-blue curves indicate  the fitted background. The inset shows the Doppler-corrected
$\gamma$-ray spectrum for all detected $p$+$^{15}$N pairs with the energies of known $\gamma$ rays indicated with the arrows. (a) Distribution gated on the $G_2$ gate shown in the inset. (b) Background-subtracted distribution gated on the $G_1$ gate in the inset. As the 2.297-MeV $\gamma$-ray sits on a significant background, the events in the $G_b$ gate, suitably scaled in magnitude,  were used to  remove this background.}   
\label{fig:p15N}
\end{figure}

\begin{figure}
  \includegraphics*[scale=.43]{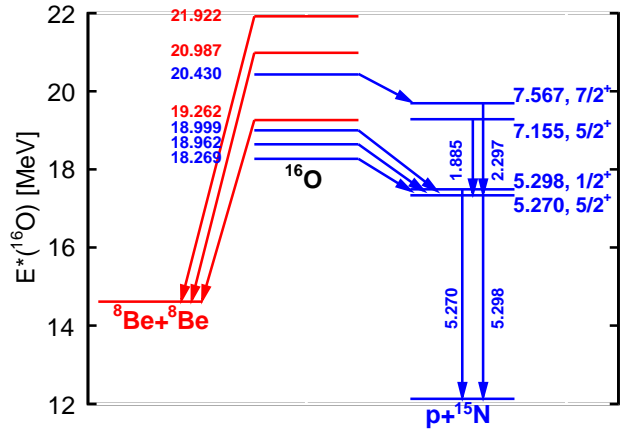}
  \caption{ Decay scheme of the newly-found  high-lying states in $^{16}$O
obtained from fits to the $p$+$^{15}$N+$\gamma$ and $^8$B$_{g.s.}$+$^8$Be$_{g.s.}$ exit channels of $^{16}$O.}

\label{fig:Oscheme}
\end{figure}

 The $\gamma$-ray spectrum measured in coincidence with the detected $p$+$^{15}$N pairs is displayed in the inset in Fig.~\ref{fig:p15N}(b). A peak at $E_{\gamma}\sim$5.28~MeV and its first escape shoulder are observed. These events can be associated with either the first ($E^*$=5.270~MeV, $J^{\pi}$=5/2$^+_1$) or 
second ($E^*$=5.298~MeV, $J^{\pi}$=1/2$^+_1$) excited state of $^{15}$N.  
In addition we see peaks at 1.885 MeV and 2.297~MeV that are produced in the decay of the $E^*$=7.155-MeV, $J^{\pi}$=5/2$^+_2$ and $E^*$=7.567-MeV, $J^{\pi}$=7/2$^{+}_1$ excited states, respectively. For reference, a partial level scheme of $^{15}$N is shown in  Fig.~\ref{fig:Oscheme}. 

The excitation-energy spectrum for events in coincidence with either the 5.270 or 5.298-MeV $\gamma$ ray (gate $G_2$ in Fig.~\ref{fig:p15N}) is plotted  in Fig.~\ref{fig:p15N}(a). Three clear peak structures are observed and the solid curve shows the results of a fit. The lower-energy peak has been fit as a doublet where the energy and width of the lower-energy member are constrained with a second $\gamma$ gate. This second gate ($G_1$) is around the 2.297-MeV $\gamma$ ray [gate $G_{1}$ in Fig.~\ref{fig:p15N}(b)] and we used the adjacent higher-energy $\gamma$ rays [gate $G_b$ in Fig.~\ref{fig:p15N}(b)] to estimate the background under this peak. The background-subtracted spectrum is displayed in Fig.~\ref{fig:p15N}(b) and only the lower-energy member of the doublet is now present as demonstrated in our fit (curve). Clearly this lower-energy member of the doublet is associated with  the 7.567-MeV, $J^{\pi}$=7/2$^+_1$ excited state of $^{15}$N  which decays by  emitting both a  2.297-MeV and a 5.270-MeV $\gamma$ ray. The deduced total excitation energies, including the $\gamma$-ray contributions are listed in Table~\ref{tbl:fit15} and the decays are illustrated in Fig.~\ref{fig:Oscheme}.

 \begin{table}
\caption{Parameters for the levels in $^{16}$O obtained from the fitting the $\gamma$-gated $p$+$^{15}$N excitation-energy distributions in Fig.~\ref{fig:p15N}. These include the fitted centroid of each
peak $E_{n\gamma}$ and its excitation energy $E^*$ when the $\gamma$-rays energies are included, and finally the fitted intrinsic width $\Gamma$.} 
\label{tbl:fit15}

\begin{ruledtabular}
\begin{tabular}{ccc}
$E^*_{n\gamma}$ & $E^*$ & $\Gamma$  \\
{[}MeV] & [MeV] &              [keV]    \\
\hline
12.863(14) & 20.430(14) & 77(38)  \\
12.993(11) & 18.269(11) & $<$30\footnotemark[1] \\
13.373(12) & 18.643(12) & $<$60\footnotemark[1] \\
13.729(12) & 18.999(12) & $<$40\footnotemark[1]

\end{tabular}
\end{ruledtabular}
\footnotetext[1]{1$\sigma$ limit}
\end{table}

\subsection{Four-$\alpha$ Exit Channels}
\begin{figure}
  \includegraphics*[scale=.43]{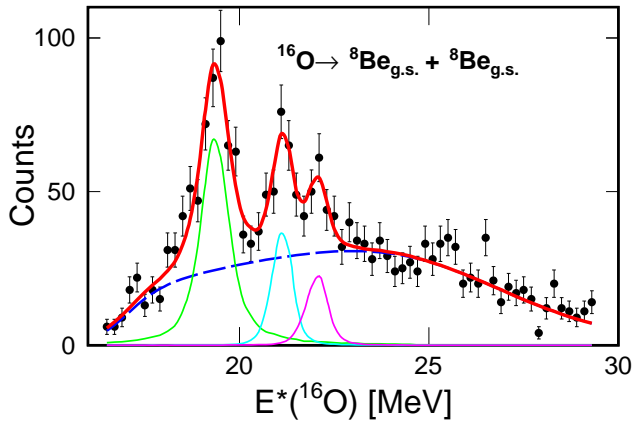}
  \caption{ Excitation-energy spectrum for the $^{8}$Be$_{g.s.}$+$^{8}$Be$_{g.s.}$ exit channel obtained from detected 4$\alpha$ events. The thick-red-solid curve show a fit to this distribution with three levels (thin solid curves)  and a smooth background (dashed-blue curve).}
\label{fig:fit88}
\end{figure}

A large number of 4$\alpha$ events were detected with the $^{15}$O beam, but the invariant-mass spectrum for all events  did not show any significant peak structures. However, such events can be obtained from a number of different decay scenarios, but one interesting possibility is the fission of $^{16}$O into two ground-state $^8$Be fragments. Such events are easy to separate by looking at the momentum correlations between the 
$\alpha$ particles. We have selected events where the relative energy between one pair of $\alpha$ particles is consistent with $^8$Be decay and similarly for the remaining pair. The relative energy distribution is very sharply peaked for $\alpha$ pairs from the decay of $^8$Be and we find there is almost no background under it. Therefore $^8$Be$_{g.s.}$+$^8$Be$_{g.s}$ decay can be isolated relatively 
cleanly.  The excitation-energy spectrum for such events is displayed in Fig.~\ref{fig:fit88} and shows a large peak at 19.26~MeV plus a broader structure at $\sim$21~MeV. The latter was fit as a doublet  in Fig.~\ref{fig:fit88} where the widths of two members were taken as equal. Fitted decay widths and cross sections are listed in Table~\ref{tbl:fit88} and the decay scheme is also illustrated in Fig.~\ref{fig:p15N}.  As the decay channel consists of two identical $J$=0 Bosons, then  these states must have positive parity and even values of $J$. Therefore they are not produced by the capture of a $d$-wave neutron, but presumably result from capture to the $f_{7/2}$ or $f_{5/2}$ levels which would not be unreasonable at these higher excitation energies. As such, these peaks must be either $J^{\pi}$=2$^+$ or 4$^+$.  
It is somewhat surprising that neutron transfer produces such  clusterized decay channels. We note it is possible that these states also have significant  proton and neutron decay branches. However we have low sensitivity to detecting such a   proton branch as it will have low efficiency and poor experimental resolution.

The $^8$Be$_{g.s.}$+$^8$Be$_{g.s.}$ exit channel of $^{16}$O has been investigated in a number of other  studies \cite{Chevallier:1967,Brochard:1976,Wuosmaa:1994,Freer:1995,Freer:2004,Soylu:2012,Curtis:2013} and a significant number of levels have been found.
Our 19.26(4)-MeV peak may be  associated with the 19.35-MeV peak originally identified by Chevallier \textit{et al.} \cite{Chevallier:1967} in the $^{12}$C($^4$He,$^8$Be)$^8$Be reaction, however, they assigned a spin of $J^{\pi}$=6$^+$ from the measured angular distributions.  Subsequently, Freer \textit{et al.} identified a peak in the $^{12}$C($^{16}$O,$^8$Be+$^8$Be)$^{12}$C reaction at 19.3 MeV and assigned a spin of $J^{\pi}$=4$^+$ \cite{Freer:2004}. Later Curtis \textit{et al.} remeasured the  $^{12}$C($^4$He,$^8$Be)$^8$Be reaction with better resolution and the 19.3-MeV peak was found to be a doublet (19.29 and 19.36~MeV) \cite{Curtis:2013}. They argued that this doublet is actually an interference effect and corresponds to a narrow resonance with either $J^{\pi}$= 2$^+$ or 4$^+$. The fitted intrinsic width of our peak is $\Gamma$=435(151);  2.9 $\sigma$ away from  zero so it is probably not narrow. In addition according to Freer \textit{et al.}, the 19.3-MeV states decays more strongly to the $\alpha$+$^{12}$C(0$^{+}_{2})$  channel with $\Gamma_{^8Be}/\Gamma_{^{12}C(0^{+}_{2})}$=0.47(15). 

\begin{figure}
  \includegraphics*[scale=.43]{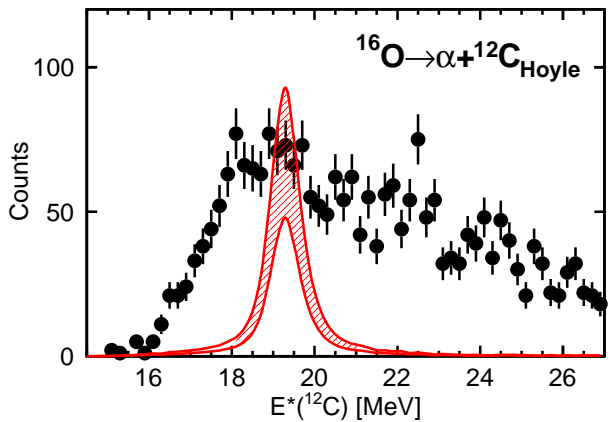}
  \caption{ Excitation-energy spectrum for the 
$\alpha$+$^{12}$C$_{Hoyle}$ exit channel obtained from detected 4$\alpha$ events. The hatched region show the simulated  range of yields from the 19.262-MeV state assuming the branching ratio and error given in \cite{Freer:1995}.}

\label{fig:hoyle}
\end{figure}

We can also relatively cleanly gate on such decays from our detected 4$\alpha$ events by selecting out those where three of the four
$\alpha$ particles has an invariant mass associated with the Hoyle state [$^{12}$C(0$^+_2$)] state. The excitation-energy spectra is displayed as the data points in Fig.~\ref{fig:hoyle}. For comparison, the two curves separated by the hatched region are simulated results using our best-fit intrinsic width for the 19.262-MeV state and  incorporating the experimental resolution. The  magnitudes of the two curves are chosen  to give the experimental outer limits of the branching ratio given by Freer \textit{et al.}  Clearly the experimental spectrum does not show such a peak and the branching strength to this channel must be at least a factor of 4 smaller than that given by Freer \textit{et al}. Probably our peak is associated with a different $^{16}$O excited state,  one that does not process pure cluster configurations but contains some neutron single-particle strength permitting its formation in neutron transfer reactions. 
In the work of Curtis \textit{et al.} \cite{Curtis:2013}, a 21.10-MeV level was observed and assigned $J^{\pi}$=4$^+$ or 6$^+$ and this 
is consistent with our  20.987(6)-MeV peak.


\begin{figure}
  \includegraphics*[scale=.43]{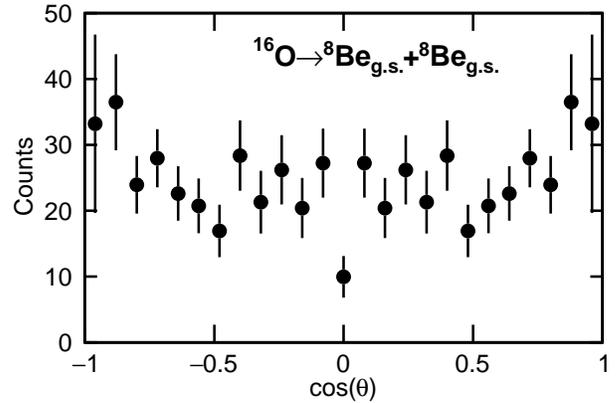}
  \caption{  Efficiency-corrected decay angular distributions for the 19.262-MeV 
state in $^{16}$O which fissions into the $^8$Be$_{g.s.}$+$^8$Be$_{g.s.}$ exit channel.} 
\label{fig:ang88}
\end{figure}

For the most significant peak at 19.262~MeV in Fig.~\ref{fig:fit88},  the angular distribution of the $^8$Be-$^8$Be axis relative to the beam direction is displayed in Fig.~\ref{fig:ang88}. It has been corrected for the angle-dependent efficiency as determined in our Monte Carlo simulations (App.~\ref{sec:MC}).  It is possible that there is some small alignment of the $^{16}$O$^*$ parent spin perpendicular to the reaction plane, but with the large error bars,  the experimental distribution is also consistent with isotropic decay. The yields quoted in Table~\ref{tbl:fit88} and subsequent tables  assumed isotropic decay in extrapolating from the transverse gate. They should only be used as a rough gauge of the cross sections unless the angular distributions are measured.

\begin{table}
\caption{Fitted mean excitation energies $E^*$, intrinsic widths $\Gamma$, 
and peak cross sections obtained for the $^{16}$O levels decaying to the $^{8}$Be$_{g.s.}$+$^{8}$Be$_{g.s.}$ channel 
observed in Fig.~\ref{fig:fit88}.} 
\label{tbl:fit88}
\begin{ruledtabular}
\begin{tabular}{ccc}
$E^*$ & $\Gamma$ & $\sigma_{peak}$ \\
{[}MeV] & [keV] & [$\mu$b]  \\
\hline
19.262(38) & 435(151)& 29(18) \\
$\left.\begin{tabular}{c}20.987(52) \footnotemark[1] \\ 21.922(87) \footnotemark[1]\end{tabular} \right\}$& 57(256)  & 14(6) \\

\end{tabular}
\end{ruledtabular}
\footnotetext[1]{doublet}
\end{table}

\section{$^{18}$Ne EXCITED STATES}
\begin{figure}
  \includegraphics*[scale=.53]{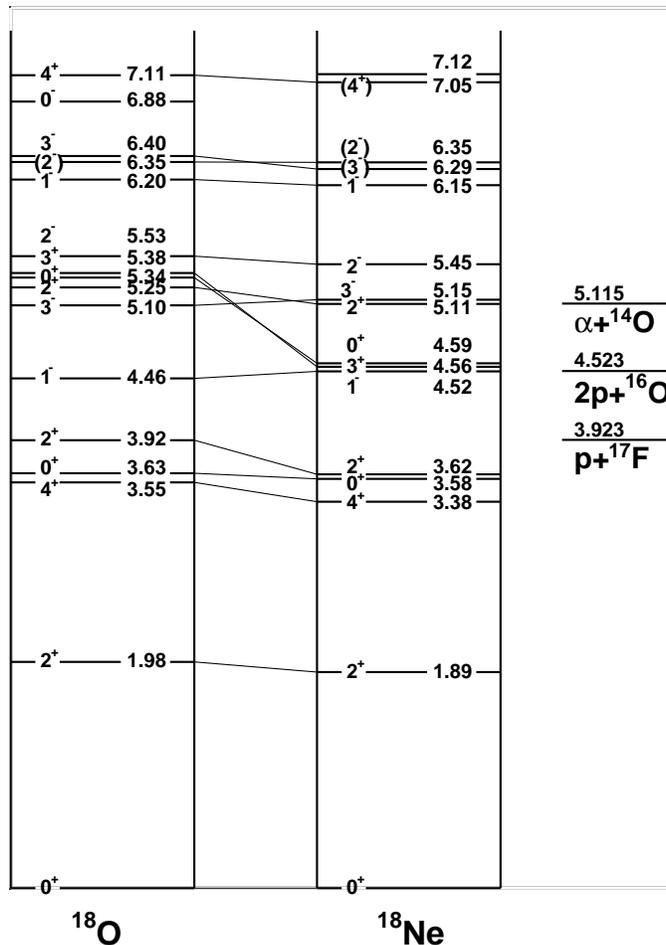}
  \caption{ Level scheme of $^{18}$Ne and and its mirror $^{18}$O as given by Hahn \textit{et al.} \cite{Hahn:1996}.}
\label{fig:scheme}
\end{figure}

 The $^{18}$Ne level scheme evaluated  by Hahn \textit{et al.} \cite{Hahn:1996} is shown in Fig.~\ref{fig:scheme} and compared to that for the $^{18}$O mirror. Some of these states can be produced by neutron capture to the $^{17}$Ne beam. The $^{17}$Ne ground-state wavefunction ($J^{\pi}$=1/2$^-_1$) consists predominantly  of two protons in the $sd$ shell, coupled to zero spin, and a single neutron hole in the $p$ shell \cite{Fortune:2006}. If the captured neutron fills in this hole, then a $J^{\pi}$=0$^+$ state in $^{18}$Ne is formed. Otherwise
neutron capture to the $sd$ shell will produce negative-parity states. Given that the momentum mismatch will favor capture to the  $d_{3/2}$ and $d_{5/2}$ levels, this reaction should predominantly populate $J^{\pi}$=1$^{-}$, 2$^{-}$, and 3$^{-}$ states. Other positive-parity states can be populated by capture to the $pf$ shell, but these will have larger excitation energies, where the level density is greater, making separation of the individual levels more difficult.

\begin{figure}
  \includegraphics*[scale=.45]{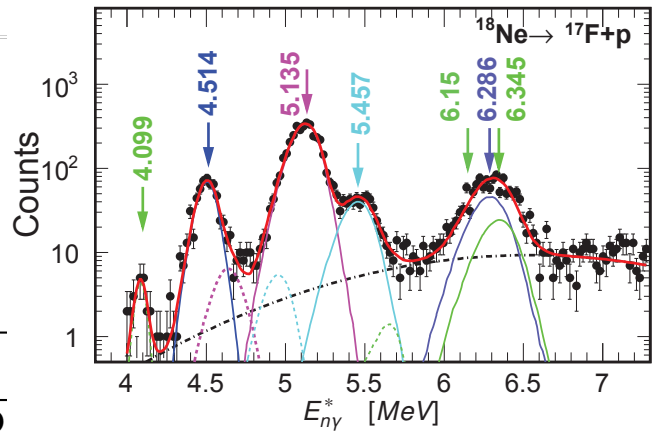}
  \caption{ Distribution of $^{18}$Ne excitation energy reduced by the total energies of emitted $^{17}$F $\gamma$-rays for  $p$+$^{17}$F pairs detected with the $^{17}$Ne beam. The experimental results are indicated by the data points. The thick-red curve shows a fit to this distribution, where individual contributions are also indicated. For each state in the fit, two peaks are included associated with decay to the ground (solid thin curves) and first-excited state (dashed thin curves) of $^{17}$F. The dot-dashed curve shows a background contribution introduced to reproduce the background in the 7-MeV region. The arrows indicate the peaks discussed in the text.}
\label{fig:fit}
\end{figure}

\begin{figure}
\includegraphics*[scale=.4]{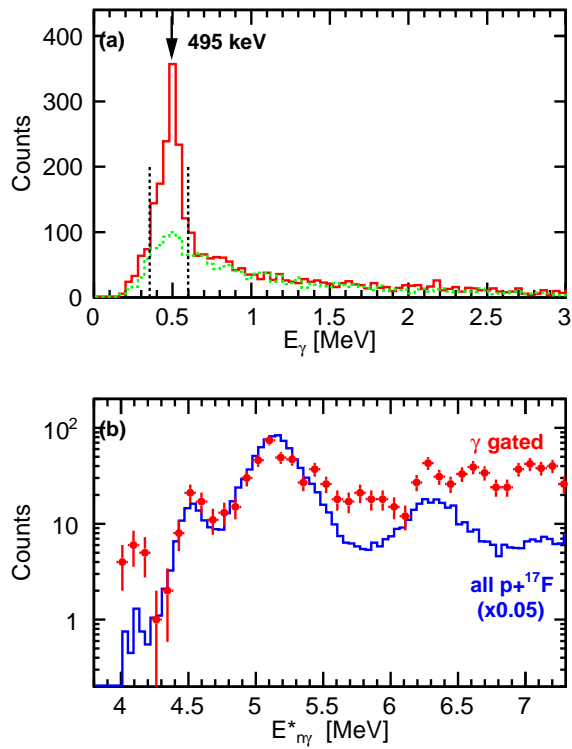}
 \caption{ (a) Spectrum of Doppler-corrected $\gamma$ rays measured with CAESAR (with add-back contributions from neighboring detectors) in coincidence with the detected $p$+$^{17}$F pairs showing the peak at 495 keV associated with the decay of the first excited state of $^{17}$F. The lower histogram shows an estimate of the background contribution, while the dashed lines indicates the outer limits of our $\gamma$-ray gate around the 495~keV peak. (b) The data points show the $\gamma$-ray-gated  spectrum of $E^*_{n\gamma}$ for detected $p$+$^{17}$F+$\gamma$ events which is compared to the histogram  for all detected $p$+$^{17}$F pairs. Both spectra were obtained with $|\cos(\theta)|<$0.7}
\label{fig:gamma}
\end{figure}

The $E^*_{n\gamma}$ distribution for transverse proton decay of $^{18}$Ne is 
shown in Fig.~\ref{fig:fit}. The residual $^{17}$F nucleus has one particle-bound excited state at 495~keV so attention must be given to the possibility of decay through this state.  The Doppler-corrected $\gamma$-ray spectrum in coincident with the $p$+$^{17}$F events is shown in Fig.~\ref{fig:gamma}(a) 
as the red-solid histogram where add-back contributions from neighboring elements are included. In comparison, the green-dashed histogram 
represents an estimate of the background under this spectrum which was  obtained from $\gamma$ rays in coincident with the prolific  2$p+^{15}$O  decay channel associated with the second excited state of $^{17}$Ne \cite{Brown:2017}. This $^{17}$Ne state does not produce $\gamma$ rays so only a 
background contribution is present. This background spectrum was normalized 
 to give the same yield  for  $E_{\gamma}> 0.8$~MeV as that for the detected 
$p$+$^{17}$F pairs. It is clear that, relative to this background, 
the $p$+$^{17}$F events have  an important contribution from the 495~keV $\gamma$-ray.

 The excitation-energy spectrum, shown as the data points in Fig.~\ref{fig:gamma}(b), is gated on the 495-keV $\gamma$ ray using the $E_{\gamma}$ limits indicated by the dashed-vertical lines in Fig.~\ref{fig:gamma}(a). It should be  compared to the inclusive spectra (blue histogram) which is normalized to the same maximum value and both were obtained requiring $|\cos\theta|<0.7$ to increase statistics. Given that there is background under the 495-keV peak, then the gated spectrum will still contain decays to the ground state of the $^{17}$F, but the decays to the excited state with be strongly enhanced.  The largest relative enhancements are found for the small $E^*\sim$4.1~MeV peak, just above the $p$+$^{17}$F threshold of 3.923~MeV, and for the background either side of  the wide $E^*\sim$6.3-MeV peak, with the enhancement of the high-energy side being largest. Therefore, these regions appear to be dominated by decay to the first excited state. The origin of the background around the 6.3~MeV peak is not clear, we do not expect very wide excited states in this region and so it must be produced from some other background process. 
           
 As the ground and first excited states of $^{17}$F are expected to have 
little neutron strength in the $sd$ shell, then the spectroscopic factor for the 
proton decay of the $^{18}$Ne states formed by neutron capture to this shell 
will be very small and hence lead to narrow intrinsic widths. The only exception would be for $J^{\pi}$=0$^+$ states formed 
by filling the neutron hole in $^{17}$Ne where larger $p$+$^{17}$F spectroscopic factors are possible. However the only observed $J^{\pi}$=0$^+$ state was close to the $p$+$^{17}$F threshold and the barrier penetration factor should also give this state a narrow width as well.  Shell-model calculations suggests the widths should be at most a few keV. In comparison our simulated
 dispersion associated with the experimental resolution has a FWHM of $\sim$200~keV. Thus in fitting the measured excitation-energy spectrum, we can ignore the contribution from the intrinsic widths and use these simulations to give the experimental line shapes. 

The fit to the excitation-energy spectrum displayed in  Fig.~\ref{fig:fit} 
was made using these line shapes and including two peaks for each level, one for decay to the ground state (solid lines) and a second peak, located 495 keV lower in mean energy, for a   decay branch to the first excited state (dashed curves). Peaks for these latter decays are not resolved in most cases, but we can 
extract maximum yields for these decays  consistent with data. The results we obtain are probably an overestimation of these excited-state branches as other sources of background are present.   In addition, there is overlap of some of these unresolved peaks and thus in the fits we consider the contributions from only one of these at a time in obtaining these limiting values. The energy, cross section and limiting branching ratio obtained from these fits are listed in Table~\ref{tbl:fit}. 

To help interpret the results we have performed shell-model calculations in 
the $spsdpf$ space with the WBP interaction \cite{Brown:2001} using the code OXBASH \cite{oxbash}. Branching ratios were calculated from the shell-model spectroscopic factors using single-particle reduced decay widths  calculated with a Coulomb plus a Wood-Saxon nuclear potential of radius parameter $r_{0}$=1.25~fm and diffuseness 0.65~fm with its depth adjusted to get the correct resonance energy.

\begin{table*}[tbp]
\caption{Parameters obtained from the fit to the  excitation-energy spectrum of $^{18}$Ne in Fig.~\ref{fig:fit}. 
The quantity  $E^*_{n\gamma}$ is the centroid of the peak in the spectrum while $E_{level}$ is the energy of the decaying level.
 These energies are different when  the decay is to the first excited state of $^{17}$F. The assigned spin-parity of the level is given by
 $J^{\pi}$, while $\sigma_{peak}$ is the cross section of the peak in the fit. Experimental and theoretical branching ratios for the decay to 
the first excited state of $^{17}$F are also listed.}

\begin{ruledtabular}
\label{tbl:fit}
\begin{tabular}{cccccc}
$E_{n\gamma}$ & $E_{level}$ & $J^{\pi}$ & $\sigma_{peak}$ & $\Gamma^*_{J^{\pi}=1/2^+}/\Gamma_{tot}$ &  $\Gamma^*_{J^{\pi}=1/2^+}/\Gamma_{tot}$ \\
{[}MeV] & [MeV] & & [$\mu$b] &exp.&theory \\
\hline
4.099(12) & 4.594(12)  & 0$^{+}_3$& 11(3)& $>$0.16\footnotemark[2] & 0.036 \\
4.514(4) & 4.514(4) & 1$^{-}_{1}$ &133(8) & $<$0.125\footnotemark[2]& 1.32$\times$10$^{-6}$ \\
5.135(2)  & 5.135(1) & 3$^-_{1}$& 1206(20)& $<$0.009\footnotemark[2] & 3.6$\times$10$^{-4}$ \\
5.457(8)  & 5.457(8) & 2$^{-}_{1}$& 186(13)& $<$ 0.19\footnotemark[2] & 0.0022 \\
 6.150\footnotemark[1]         & 6.150\footnotemark[1]    & 1$^{-}_{2}$ & $<$54\footnotemark[2]&0.65\footnotemark[3] \\
$\sim$6.3  & $\sim$6.3  & (2$^{-}_{2}$,3$^{-}_{2}$)& 354(17)& $<$0.12\footnotemark[2] & \\
\end{tabular}
\end{ruledtabular}
\footnotetext[1]{Fixed to value from \cite{ENSDF}}
\footnotetext[2]{2$\sigma$ limit}
\footnotetext[3]{Fixed to value from \cite{Blackmon:2003}}

\end{table*}

\subsection{4.099-MeV Peak}
\label{sec:lowest}
The lowest-energy peak observed in Fig.~\ref{fig:fit} is about 200~keV above the 
3.923-MeV threshold for the $p$+$^{19}$F decay channel. From Fig.~\ref{fig:gamma}, 
we argued that this peak is associated with decay to the first excited state of $^{17}$F rather than the ground state like the other observed  peaks. Given that the decay energy to the ground state is much larger ($\sim$700~keV above threshold) one might expect its smaller barrier penetration factor would kill any significant decay branch to the excited state unless this state had some special structure. 

Including the $\gamma$-ray energy (495 keV), our peak corresponds to a level at  $E^*$=4.594(12)~MeV which is consistent with the energy of the $J^{\pi}$=0$^{+}_{3}$ level measured by Nero \textit{et al.} (see Sec.~\ref{sec:4p5}). The structure of the lowest three 0$^+$ states in $^{18}$Ne can be gauged from studies of their analogs in $^{18}$O. Fortune and Hadley argue that these states have proton $(1s_{1/2})^2$ and $(0d_{5/2})^2$ components as well as a collective 4p-2h contribution \cite{Fortune:1974}.  They also indicate that the wavefunction for the third of these states is dominated by the $(1s_{1/2})^2$ contribution which will give a large spectroscopic factor for the $p$+$^{17}$F$^{*}_{J=1/2^+}$ decay channel. Of course the $(0d_{5/2})^2$ component will be associated with decay to the $J^{\pi}$=5/2$^+$ ground state of $^{17}$F. In addition to the larger spectroscopic factor for decay to the excited state, this mode will be further enhanced by a smaller centrifugal barrier; $\ell$=0 compared to $\ell$=2 for ground-state decay.  Both of these two properties conspire to counter the effect of the small decay energy and  give a significant branch to the excited state. However we expect that decay to the ground state is also significant. Yield from such a branch would produce an enhancement to the high-energy tail of the 4.514-MeV peak (Sec.~\ref{sec:4p5}). With the maximum amount of this contribution allowed in our fit, we conclude that the minimum branching ratio to the first excited state is 16\% at the 2$\sigma$ level.

Our shell-model predictions give a value of 3.6\% for this branching ratio 
using the level energy 4.950(8)~MeV listed in \cite{ENSDF}. The calculated branching  ratio is quite sensitive to this energy, with its value increasing to 7.6\% if the energy is increased by twice its statistical uncertainty.
However it is still smaller than the experimental lower limit of 16\% suggesting that the relative contribution of 
$(1s_{1/2})^2$ to $(0d_{5/2})^2$ of 5.5 is underestimated in these shell-model calculations. In the work of Fortune and Hadley, the strengths of the different configurations in the 0$^+$ wavefunctions  were constrained using 
experimental data giving a  $(1s_{1/2})^2$ to $(0d_{5/2})^2$ ratio of 14.4 for this state.
This is a factor of 2.6 larger than our shell-model calculations and 
allows for consistency with our  experimental limit.

The shell model predicts a  large spectroscopic factor of $C^2S(p_{1/2}$)=0.66 for neutron capture to the $p_{1/2}$ level. However the larger momentum mismatch for $p$-wave capture should suppress the yield of this case  relative to those for $d$-wave capture. We measured a cross section of 13(3)~$\mu$b for the proton decay branch to the first excited state of $^{17}$F. However, based on the minimum limit for this branching ratio in Table~\ref{tbl:fit}, the total cross section for this state must be less than 81$\mu$b. This is more than a factor of 15 smaller than the yield for the 5.135-MeV, $J^{\pi}$=3$^-_1$ state (Sec.~\ref{sec:5p135}) which has a predicted spectroscopic factor of similar  magnitude, but is associated with $d$-wave capture. This result is thus consistent with a large suppression due to the momentum mismatch. 

\subsection{4.514-MeV Peak}
\label{sec:4p5}
 Nero \textit{et al.} \cite{Nero:1981} reported a 
doublet at $E^*\sim$4.5~MeV. In the $^{16}$O($^3$He,$n$)$^{18}$Ne reaction the level energies were determined as 4.513(13) and 4.587(13)~MeV  while in the $^{20}$Ne($p$,$t$)$^{18}$Ne reaction they are 4.522(10) and 4.592(10)~MeV, respectively. Nero \textit{et al.} concluded that the lower-energy member is $J^{\pi}$=1$^-_1$ while  the higher-energy member is $J^{\pi}$= 0$^{+}_{3}$. Our peak at 
$E^*$=4.514(4)~MeV is thus consistent with the $J^{\pi}$=1$^-_{1}$ level.

Although we list a limit of 12.5\% for the excited-state branching ratio, the actually value is expected to be extremely small
as decay to the excited state is only 97~keV above threshold compared to 592~keV for ground-state decay. The shell-model estimate is 
$~\sim$10$^{-6}$.

The $n$+$^{17}$Ne spectroscopic factor predicted for this state is large, however the shell-model calculations suggested it should be largely due to $s$-wave capture [$C^2S$($d_{3/2}$)=0.015, $C^2S$($s_{1/2}$)=0.365] and thus should be suppressed due to the larger momentum mismatch. Either the effect of the momentum mismatch is not as large as we expect or these shell-model predictions are in error.

\subsection{ 5.135-MeV Peak}
\label{sec:5p135}
The dominant peak in the excitation-energy spectrum of Fig.~\ref{fig:fit}
 occurs at 5.135(2) MeV. Nero \textit{et al.} \cite{Nero:1981} reported on  a 
doublet at $E^*\sim$5.1~MeV using data from two reactions.
In the $^{16}$O($^{3}$He,$n$)$^{18}$Ne reaction, the level energies were 
determined as 5.075(13) and 5.135(25)~MeV, while in the 
$^{20}$Ne($p$,$t$)$^{18}$Ne reaction they are 5.099(10) and 5.151(10)~MeV.  
From angular distributions measured in that work and also by 
Falk \textit{et al.} \cite{Falk:1970}, one of these states was determined to be a 
$J^{\pi}$=2$^{+}_3$ and the other a $J^{\pi}$=3$^-_1$, but which one is the 2$^{+}_3$, and conversely,
 which one is the 3$^-_1$ was unknown.  

In order to reproduce the measured intrinsic widths of these states, Hahn \textit{et al.} 
\cite{Hahn:1996} subsequently argued that the higher-energy state is 
$J^{\pi}$=3$^-_1$, 
while the lower-energy state is $J^{\pi}$=2$^{+}_3$. This is in contrast to Wiescher \textit{et al.}  \cite{Wiescher:1987} and  Funck \textit{et al.} \cite{Funck:1988,Funck:1989} 
who put these states in reverse order in their $^{14}$O($\alpha$,$p$)$^{17}$F rate calculations for astrophysics.


If these two peaks were both present in our data, our energy resolution would not be sufficient to separate them, however given that this reaction is not expected to excite the $J^{\pi}$=2$^{+}$ level significantly, we conclude that the peak observed at $E^*$=5.135(2)~MeV is associated predominantly with the $J^{\pi}$=3$^-_1$ state.
With our $\pm$6.6~keV systematic uncertainty (Sec.~\ref{sec:method}), its energy is consistent with only the higher-energy member of the doublet as measured by Nero \textit{et al.} and thus with the spin order given by Hahn \textit{et al.} In the shell-model calculations, this state has the largest spectroscopic factor for neutron capture to a $d$ level [$C^2S$=0.65 ($d_{5/2}$)] and therefore it is not surprising that it is the strongest state populated in this reaction.

Almaraz-Calderon \textit{et al.} observed a peak at a similar energy ($E^*$=5.10(10) MeV) in the $^{16}$O($^3$He,$n$) reaction but did not have enough resolution to separate the two members of the doublet if they both were present.  
They measured a branching ratio to the first excited state of $^{17}$F of 
0.110 which is large compared to our upper limit of 0.009. The $J^{\pi}$=2$^+_3$ member would have to have a large branching ratio and contributed significantly to their observed peak to be consistent with our results.  However, our shell-model calculations suggest that this 2$^+$ state has a very small branching ratio of 0.002.

\subsection{5.457-MeV peak}
 A state is resolved on the higher-energy side of the dominant 5.135-MeV peak in Fig.~\ref{fig:fit} at 5.457(8)~MeV.
 This energy is consistent with a level at 5.453(10)~MeV measured by Nero \textit{et al.} 
in the $^{20}$Ne($p$,$t$) reaction \cite{Nero:1981}. However, no other information on this level was determined  due to its low population in that work. Hahn \textit{et al.} list a level at 5.454 MeV as $J^{\pi}$=2$^{-}_1$ based on Coulomb-energy shifts and  angular distributions in two transfer reactions, but mostly the fact that  the analogs of all other $^{18}$O  excited states in this energy region have all been identified except for this $J^{\pi}$=2$^-_1$ state. The observation of a 5.457(8)-MeV state in this work confirms this assignment. The shell-model calculations also suggest that this state has a strong $n$+$^{17}$Ne spectroscopic factor
 with $C^2S(d_{5/2})$=0.23 and  $C^2S(d_{3/2})$=0.12.

\subsection{6.3-MeV Peak}
The second-most intense peak seen in Fig.~\ref{fig:fit} occurs at approximately 6.3~MeV with a  width that is  larger than the predicted experimental resolution for this energy. Assuming that the intrinsic widths of all states in this region as very small, then this peak must be a multiplet.
Hahn \textit{et al.} list three negative-parity levels in this energy region that could be excited in our reaction \cite{Hahn:1996};  a $J^{\pi}$=3$^-$, 2$^-$ doublet at $E^*$=6.286 and 6.345~MeV and in addition the   $J^{\pi}$=1$^-$ level at $E^*$=6.15~MeV that can contribute to the low-energy tail.
For this latter state,  He \textit{et al.} determine that the excited-state and ground-state decay branches are approximately 
equal \cite{He:2009}, while Blackmon \textit{et al.} measured $\Gamma^*_{J=1/2^{+}}/\Gamma_{tot}$ = 0.73 \cite{Blackmon:2003}. In addition our shell-model calculations also give a large branching ratio, $\Gamma^*_{J=1/2^+}/\Gamma_{tot}$=0.65. With such values, any ground-state-decay yield that makes a significant contribution to the low-energy side of the  6.3-MeV peak will produce too much yield in the $E^*$ region associated with excited-state decay. Thus we conclude that this level does not contribute significantly to the observed peak.

The fit shown in Fig.~\ref{fig:fit} was obtained as the sum of two peaks of similar intensities with energies of 6.279(36) and 6.369(36) MeV which are consistent with the energies of the aforementioned doublet listed by Hahn \textit{et al.}  The spin order of this doublet is not well determined, but the preference of Hahn \textit{et al.} is the opposite order to that for the analog states in $^{18}$O (see Fig.~\ref{fig:scheme}). In Table~\ref{tbl:fit} we list only only the total cross section and the average branching ratio for these two states.



\subsection{Decay Angular Distributions}
\label{sec:ang}
In our Monte Carlo simulations (App.~\ref{sec:MC}) used in the fitting of the excitation 
spectrum for  $|\cos\theta|<0.2$, we have assumed that the decay of the $^{18}$Ne fragments are isotropic in space. However for these transfer reactions one should consider the possibility that the spin vectors of the $^{18}$Ne states have a strong 
alignment perpendicular to the reaction plane leading to deviations from isotropic emission.  As such, the extrapolation from  the  $|\cos\theta|<0.2$ region will be incorrect leading to errors in the extracted cross sections and branching ratios presented in  Table~\ref{tbl:fit}.

\begin{figure}
  \includegraphics*[scale=.43]{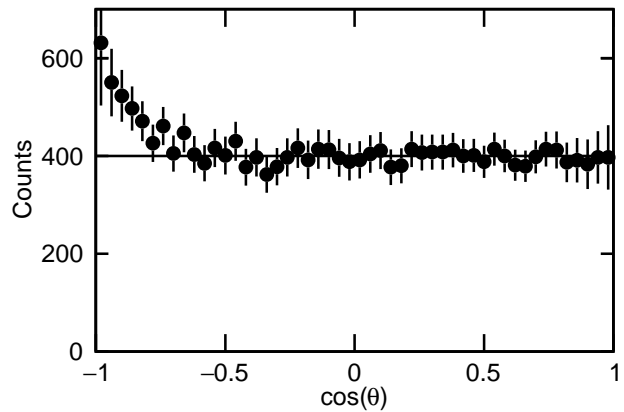}
  \caption{ Experimental  angular distributions for the decay of the 
5.135-MeV, $J^{\pi}$=3$^-_1$ state in $^{18}$Ne corrected for the detector efficiency.}
\label{fig:ang}
\end{figure}

The best state to look for such an effect is the dominant 5.135-MeV state where the statistics are large and the background is small apart from a $\sim$10\% contribution from the neighboring 5.457-MeV state which cannot be separated from the dominant peak at larger  $|\cos\theta|$ values due to the degraded resolution. The efficiency-corrected $\cos\theta$ distribution is plotted as the data points in Fig.~\ref{fig:ang}. This distribution is largely isotropic apart from an enhancement at $\cos\theta\sim$-1. As the distribution should be symmetric about $\cos\theta$=0, this enhancement cannot be real and may be associated with the background. 

While a similar analysis is not possible for the other states due to statistical and background issues, we find that simulations of 
$p_{3/2}$ decay of aligned ($M$=0 projection on beam axis) $^{18}$Ne fragments to $^{17}$F$_{g.s.}$ 
can only lead to, at most, a reduction of 30\% in yield due to the  extrapolation to larger $|\cos\theta|$ values. On the other hand in the decay to the excited state of $^{17}$F, we find instead enhancements in the yield due to this extrapolation  of up to a factor of 2 for $p_{3/2}$ and $f_{7/2}$ decays. If there is significant alignment, then our limits to the branching ratios in Table~\ref{tbl:fit} obtained from the isotropic simulations will be too small for $J\neq0$. 
However based in result in Fig.~\ref{fig:ang}, we do not expect this to be significant.

\subsection{Branching ratios}
The extracted limits to the branching ratios to the first excited state of $^{17}$F  are listed in Table~\ref{tbl:fit} and compared to values from our shell-model calculations. Some of these cases have already been discussed in the previous sections. Apart from the 4.594-MeV $J^{\pi}$=0$^+_3$ state,  our maximum limits are all much larger than, and thus consistent with, the theoretical values.  The only other negative-parity state which is expected to have a significant branching ratio, the 6.150~MeV $J^{\pi}$=1$^-_2$ level \cite{He:2009,Blackmon:2003}, was not resolved in this work but may contributed to the enhanced yield of the $\gamma$-ray gated yield in Fig.~\ref{fig:gamma}(b) between the 5.349 and 6.3-MeV peaks. 

\subsection{Other exit channels}
\begin{figure}
  \includegraphics*[scale=.43]{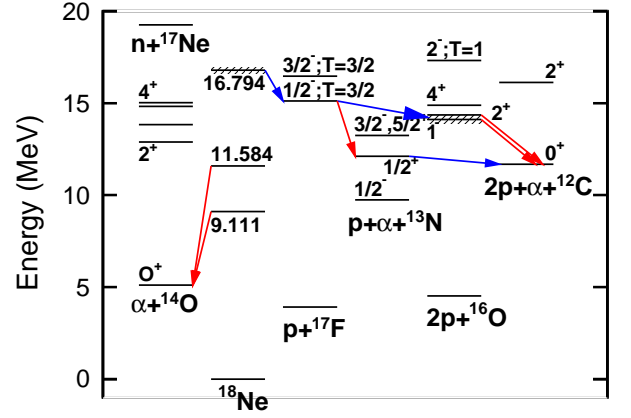}
  \caption{Level diagram for $^{18}$Ne and the neighboring nuclei showing the decay of the three new states associated with the $\alpha$+$^{14}$O and 2$p$+$\alpha$+$^{12}$C channels. }

\label{fig:levela14O}
\end{figure}

Apart from the $p$+$^{17}$F exit channel, we have also observed three peaks in  the $\alpha$+$^{14}$O and 2$p$+$\alpha$+$^{12}$C invariant-mass distributions which correspond to higher-lying excited states. The extracted level information is listed in Table~\ref{tbl:a14O} and the decay of the states are illustrated in the level diagram in Fig.~\ref{fig:levela14O}. No evidence of these levels has been observed in other decay channels, though the $p$+$^{17}$F decay channel in particular will have low efficiency and poor resolution so our sensitivity is significantly reduced.

\begin{figure}
  \includegraphics*[scale=.43]{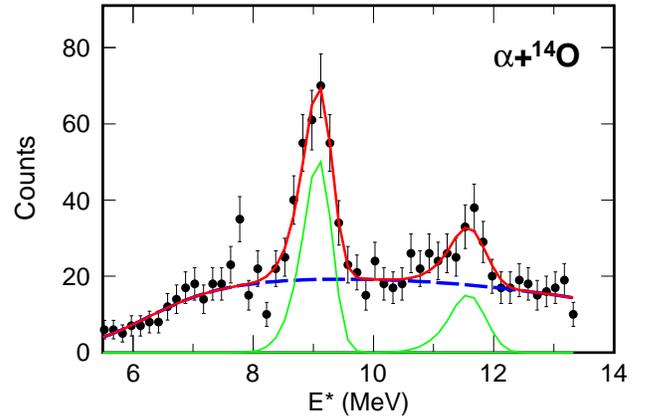}
  \caption{Experimental $^{18}$Ne excitation-energy distribution for transverse $\alpha$+$^{14}$O decays. The solid red curve shows a fit to this data with the smooth fitted background shown as the dashed-blue curve and the individual peaks as the solid green curves. }

\label{fig:Exa14O}
\end{figure}

The excitation-energy distribution from the $\alpha$+$^{14}$O channel is shown in Fig.~\ref{fig:Exa14O}. A rather narrow level ($\Gamma <$ 60~keV) is observed at 9.111(25)~MeV and a higher-energy peak  is also present at 11.58(64)~MeV.  
 The lower-energy peak was not observed in an $\alpha$+$^{14}$O elastic scattering experiment, where a $E^*$=9.2~MeV level was found, but its width is much larger ($\Gamma$=300~keV) \cite{Fu:2008}. The presence of the wider peak at almost the same energy may have reduced their sensitivity to the level 
we observed, but on the other hand with its small decay width, it may not have a strong $\alpha$-cluster structure and thus was not strongly excited in the $\alpha$-scattering experiment.

\begin{table}
\caption{Fitted mean excitation energies $E^*$, intrinsic widths $\Gamma$, 
and cross sections of states obtained from fitting the $^{18}$Ne$\rightarrow \alpha$+$^{14}$O 
decay spectrum  in Fig.~\ref{fig:Exa14O} and the $^{18}$Ne$\rightarrow$2$p$+$\alpha$+$^{12}$C decay spectrum in Fig.~\ref{fig:ppa12C}(a).} 
\label{tbl:a14O}
\begin{ruledtabular}
\begin{tabular}{cccc}
$E^*$ & channel& $\Gamma$ & $\sigma_{peak}$ \\
{[}MeV] &       & [keV] & [$\mu$b]  \\
\hline
9.111 (25)   &   $\alpha$+$^{14}$O &  $<$60\footnotemark[1] & 52(5) \\    
11.584 (64)  &   $\alpha$+$^{14}$O &  $<$ 650\footnotemark[1] & $\sim$18\\
16.794(29) & 2$p$+$\alpha$+$^{12}$C & 328(68) & 182(11)\\
\end{tabular}
\end{ruledtabular}
\footnotetext[1]{1$\sigma$ limit}
\end{table}
 
\begin{figure*}
  \includegraphics*[scale=.8]{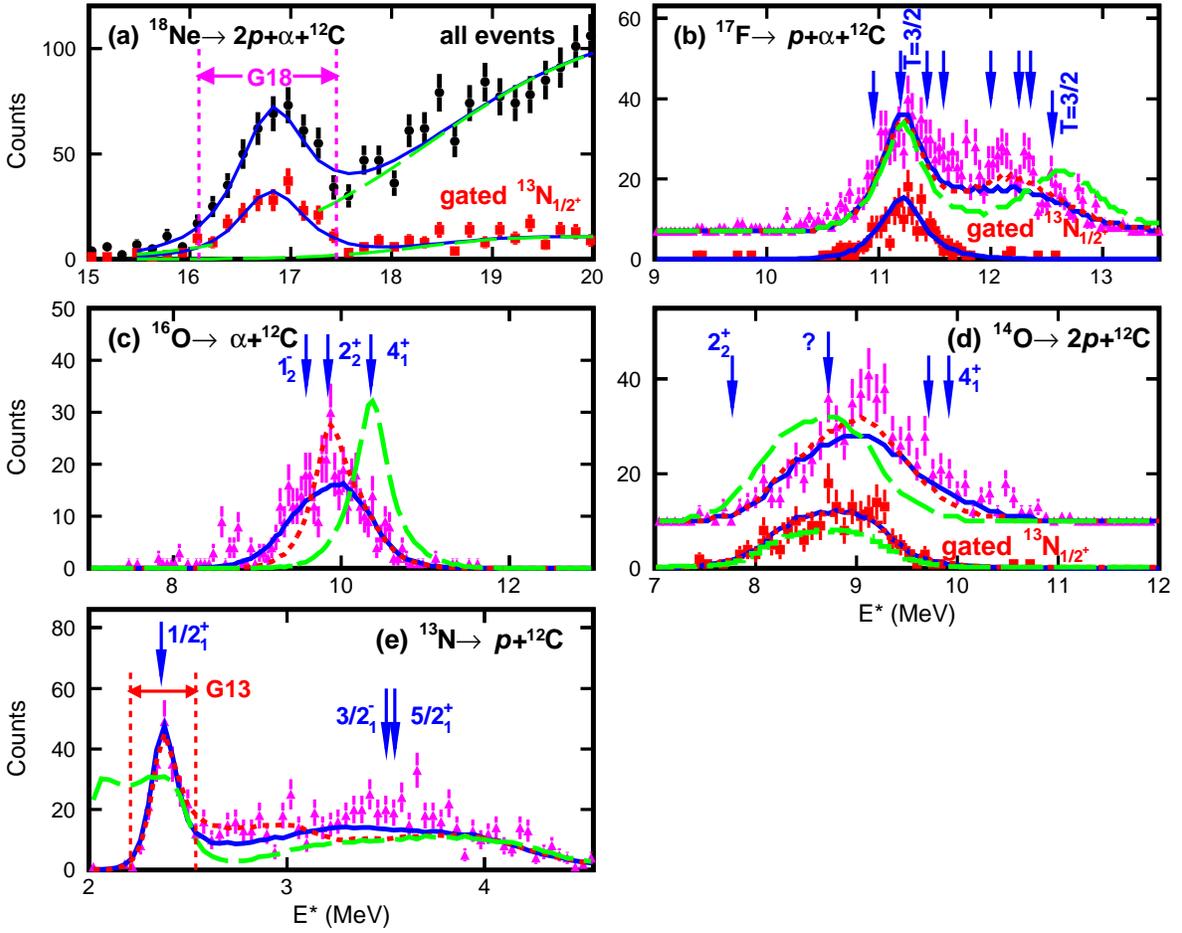}
  \caption{Experimental invariant-mass spectra obtained from 2$p$+$\alpha$+$^{12}$C events. (a) The black circular data points show the $^{18}$Ne excitation energy distribution for all events (no transverse gate) while the red square data have a gate requiring an $^{13}$N$_{1/2^{+}_{1}}$ intermediate state was present. The solid blue curves are fits to the data where the fitted smooth backgrounds are shown by the dashed green curves.  (b-e), The magenta triangular data are invariant-mass distributions of possible intermediate states gated on the observed $^{18}$Ne peak [gate $G18$ in (a)]. Arrows mark the location of states listed in the ENSDF data base \cite{ENSDF}. The data in (b) and (d) have been shifted along the $y$ axis for clarity. The red-square data in these two panels have an extra gate [gate $G13$ in (e)] applied.  The solid blue, red dotted, and green dashed curves are predictions from a simulation where  the IAS in $^{17}$F has a second decay branch to either the $J^{\pi}$=1$^-_2$, 2$^+_2$, or 4$^+_1$ excited state in $^{16}$O.}

\label{fig:ppa12C}
\end{figure*}

For highly-fragmented  decay channels,  it can be difficult to determine the decay path as there are many possible 
intermediate states  and it become especially difficult if there are multiple decay paths  as is the case for the peak in 2$p$+$\alpha$+$^{12}$C channel. The invariant-mass spectrum for this channel, shown as the  black circular data points in Fig.~\ref{fig:ppa12C}(a), contains a peak at 16.794(20) MeV. 
Due to the low statistics, no transverse gate has been  applied for this channel.
After selecting events in this peak [gate $G18$ in Fig.~\ref{fig:ppa12C}(a)], the excitation-energy spectra of the various possible intermediate states are plotted in Figs.~\ref{fig:ppa12C}(b) to \ref{fig:ppa12C}(e) as the magenta triangular data points. As there are two possible protons to construct the potential $^{17}$F$\rightarrow$ $p$ + $\alpha$+ $^{12}$C and $^{13}$N$\rightarrow p$ + $^{12}$C intermediate states, we have determined the excitation energy using each of these protons in turn, i.e., these spectra were incremented twice for each event. For comparison, the arrows show the locations of the energy levels listed in the ENSDF data base \cite{ENSDF}. Of the possible intermediate states, one stands out very clearly, the  1/2$^{+}_{1}$, first excited state of $^{13}$N at $E^*$=2.365 in Fig.~\ref{fig:ppa12C}(e). To confirm this state is associated with the peak and not the $\sim$30\% background under the peak, we have gated on the $^{13}$N peak [gate $G13$ in Fig.~\ref{fig:ppa12C}(e)] and the corresponding $^{18}$Ne spectrum is shown as the red square data points in Fig.~\ref{fig:ppa12C}(a). The fitted yield in this new gated $^{18}$Ne spectrum is about half of the ungated version if smooth backgrounds (dashed curves) are assumed in fits.  Thus we conclude  that the $^{18}$Ne level has at least two decay pathways, one of which decays in a manner that produces the 1/2$^{+}_1$ $^{13}$N intermediate state and one that does not.

Let us concentrate of the decay pathway though the $J^{\pi}$=1/2$^+_1$, $^{13}$N state first. If the $^{18}$Ne state  decays via a series of sequential decay steps, then in order to pass through the $^{13}$N intermediate state, it must first  decay to a $^{17}$F or $^{14}$O intermediate state. See the level schemes of these and other nuclei of interest in Fig.~\ref{fig:levela14O}. To search for such states, we have further applied the $G13$ gate on the $^{17}$F and $^{14}$O excitation-energy spectra in Figs.~\ref{fig:ppa12C}(b) and \ref{fig:ppa12C}(d) (red square data points).  For the $^{17}$F case, this gated yield is peaked around the energy of the known isobaric analog state (IAS) ($T$=3/2, $J^{\pi}$=1/2$^-$, $\Gamma$=0.18 keV) at $E^*$=11.192~MeV. The solid curve through these data points is a simulation of the detector response of this narrow state which reproduces its shape very well.  Thus we conclude that this decay pathway is  described by an initial proton decay to the $^{17}$F$_{IAS}$ which subsequently $\alpha$ decays to the $^{13}$N state, which then proton decays to the ground state of $^{12}$C.

Given that this new $^{18}$Ne state has a strong  proton-decay branch to a high-$T$ state in $^{17}$F, it is quite probable that this new $^{18}$Ne state is itself high $T$, i.e, $T$=2 in this case. Its excitation energy is appropriate for it to be an analog of a low-lying state in $^{18}$Na (see later).  Now if the second decay pathways involves a second decay branch of $^{18}$Ne, then to conserve isospin and energy, it should be a proton decay to the next analog state in $^{17}$F at $E^*$=12.550~MeV. However, the latter decay is only $\sim$300~keV above threshold and will be suppressed by the small Coulomb penetration factor. In addition we do not see any indication of significant yield for this intermediate  state in Fig.~\ref{fig:ppa12C}(b).  Thus it is more likely that the second decay pathway involves a second decay branch of $^{17}$F$_{IAS}$. Note that $^{17}$F$_{IAS}$ itself, has no isospin-allowed particle decay modes which are above threshold, so we expect all of its decay branches to violate isospin symmetry.   

We have dismissed the possibility that this  second decay branch of $^{17}$F$_{IAS}$ is an $\alpha$-decay to higher-lying states of $^{13}$N as  there is no indication of any significant yield for such states in Fig.~\ref{fig:ppa12C}(e). Thus we restrict ourselves to a proton decay branch to either the 1$^{-}_{2}$, 2$^{+}_{2}$, or 4$^{+}_{1}$ excited state in $^{16}$O. 
As such, we have simulated the decay of the $^{18}$Ne state as an initial proton decay to $^{17}$F$_{IAS}$, followed by either another proton decay to one of these three $^{16}$O intermediate states or alternatively an $\alpha$ decay to the $J^{\pi}$=1/2$^{+}_1$, $^{13}$N intermediate state, with these latter intermediate states subsequently decaying to give us the 2$p$+$\alpha$+$^{12}$C exit channel. For each possible $^{16}$O intermediate, the $p$/$\alpha$ branching ratio of $^{17}$F$_{IAS}$ was adjusted to best fit both the gated and ungated $^{18}$Ne excitation-energy spectra in Fig.~\ref{fig:ppa12C}(a). The simulated $^{17}$F, $^{16}$O, $^{14}$O, and $^{13}$N invariant-mass spectra are then compared to the experimental data in Figs.~\ref{fig:ppa12C}(b) to \ref{fig:ppa12C}(e) as the solid, dotted and dash curves respectively.  As there is roughly a 30\% background under the ungated  $^{18}$Ne peak in Fig.~\ref{fig:ppa12C}(a), the predicted distributions should not account for the total experimental yield in these panels. 
Thus consistency with the experiment data occurs if these simulated distributions do not pass above the data points. In this regard, the simulation for the 4$^{+}_{1}$ $^{16}$O intermediate state (green dashed curves) must be clearly be discarded.  The simulation for the 1$^{-}_{2}$ state (solid blue curves) is consistent with all distributions, while for the 2$^{+}_{2}$ state (magenta dotted curve),  the curve in Fig.~\ref{fig:ppa12C}(c) overshoots the experiments distribution by roughly 30-50\% at its peak. Thus the second decay branch of $^{17}$F$_{IAS}$ involves proton decay to the 1$^{-}_{2}$ $^{16}$O state, but we cannot rule out in addition some smaller branch to the 2$^{+}_{2}$ state and smaller yields for other decay paths. The fitted branching ratio of $^{17}$F$_{IAS}$ is $\Gamma_{\alpha}/\Gamma_{p}$=0.65(9).

\begin{figure}
  \includegraphics*[scale=.44]{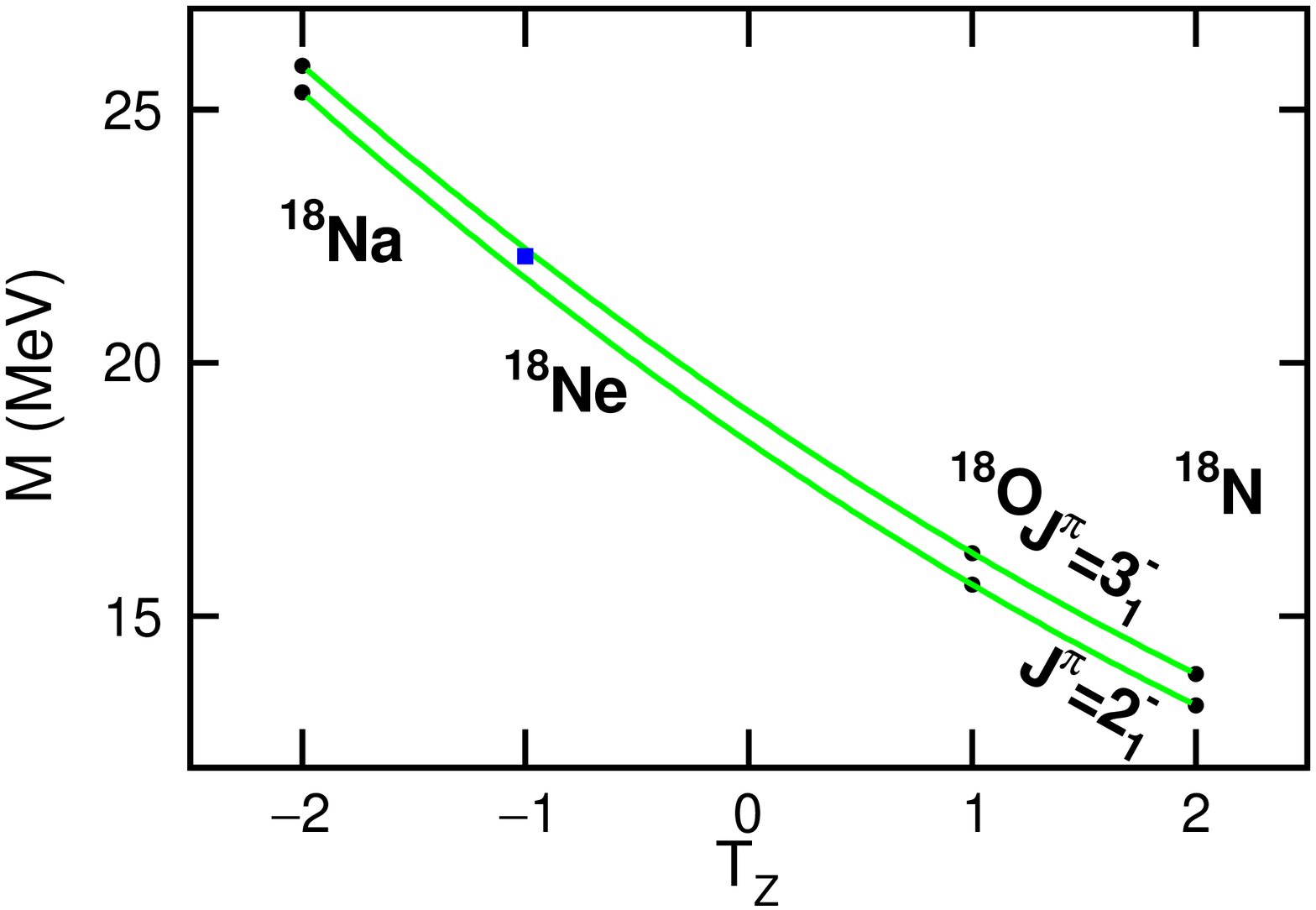}
\caption{ Known mass excesses of the $J^{\pi}$=2$^-_1$ and 3$^-_1$, $A$=8, $T$=2 multiplets are plotted as the circular data points. The curves are fits with the IMME [Eq.~(\ref{eq:IMME})]. The location of the $^{18}$Ne$\rightarrow$ 2$p$+$\alpha$+$^{12}$C state is shown by the blue square.}

\label{fig:IMME}
\end{figure}

For an isospin multiplet, the mass excesses are expected to be well described by the isospin multiplet mass equation (IMME) \cite{MacCormick:2014}
\begin{equation}     
M(T,T_{Z}) = a + b T_{Z} + c T_{Z}^2.
\label{eq:IMME}
\end{equation}
where $a$, $b$, and $c$ are constants. Except for a few cases, deviations from the quadratic $T_{Z}$ dependence are quite small. For the $A$=18, $T$=2 multiplets, only a few cases have at least three members known to constrain the three constants. In Fig.~\ref{fig:IMME} we show quadratic IMME  fits to the $J^{\pi}$=2$^{-}_{1}$ and 3$^{-}_{1}$ members using mass excesses determined for $^{18}$Na from  \cite{Assie:2012}. For the  $^{18}$O, and $^{18}$N cases, we have used ground-state masses from the AME2016 tabulation \cite{AME2016} and excitation energies from \cite{ENSDF,Hoffman:2013}. For comparison, the location of the new $^{18}$Ne peak is shown as the blue square data point. It is closer to the fitted curve for $J^{\pi}$=3$^{-}_{1}$ levels, but 140(34)~keV below. Generally we expect deviation from the IMME to be much smaller than this, so probably the observed peak in not purely from this level in $^{18}$Ne. Indeed the fitted intrinsic width of this state is relatively large, $\Gamma$=328(68) keV, significantly larger than that of the 3$^{-}$ state in $^{18}$Na ($\Gamma$=42(10) keV \cite{Assie:2012}). In $^{18}$Na, a very wide [$\Gamma$=900(100)~keV] was observed $\sim$50~keV below this 3$^{-}_{1}$ state while a very narrow state ($\Gamma < $1~keV) was observed $\sim$100 keV below. It is  possible that the observed peak is a multiplet with contributions from a number of $^{18}$Ne levels in this energy region. 

\section{$^{10}$C EXCITED STATES}
The ground state of $^{9}$C is $J^{\pi}$=3/2$^-$. This is mostly a $p$-shell nucleus and the transfer of another neutron into the $p$-shell will populate 
$J^{\pi}$=0$^+$, 1$^{+}$, 2$^{+}$, and 3$^+$ states in $^{10}$C. At higher excitation energies,  negative-parity levels can be populated by adding the extra neutron to the $sd$ shell.

The ground and first excited states of $^{10}$C are particle bound and at $E^*$= 3.73~MeV, the 2$p$+2$\alpha$ decay channel opens up. This is the only available final exit channel for particle decay until $E^*$=15.0~MeV when the $^{3}$He+$^7$Be channel is available. A number of invariant-mass studies have investigated  2$p$+2$\alpha$ exit channels produced in the inelastic excitation of the a $^{10}$C beam \cite{Charity:2007, Mercurio:2008,Curtis:2008,Charity:2008}.
Numerous states were observed whose decay are initiated by either by $p$, $\alpha$, or direct two-proton emission. In all the cases, the remnant nucleus  undergoes further particle emission producing the observed exit channel. Many of the states are expected to have large $\alpha$-particle cluster structure like that of the 
ground-state configuration.

\begin{figure*}
  \includegraphics*[scale=.86]{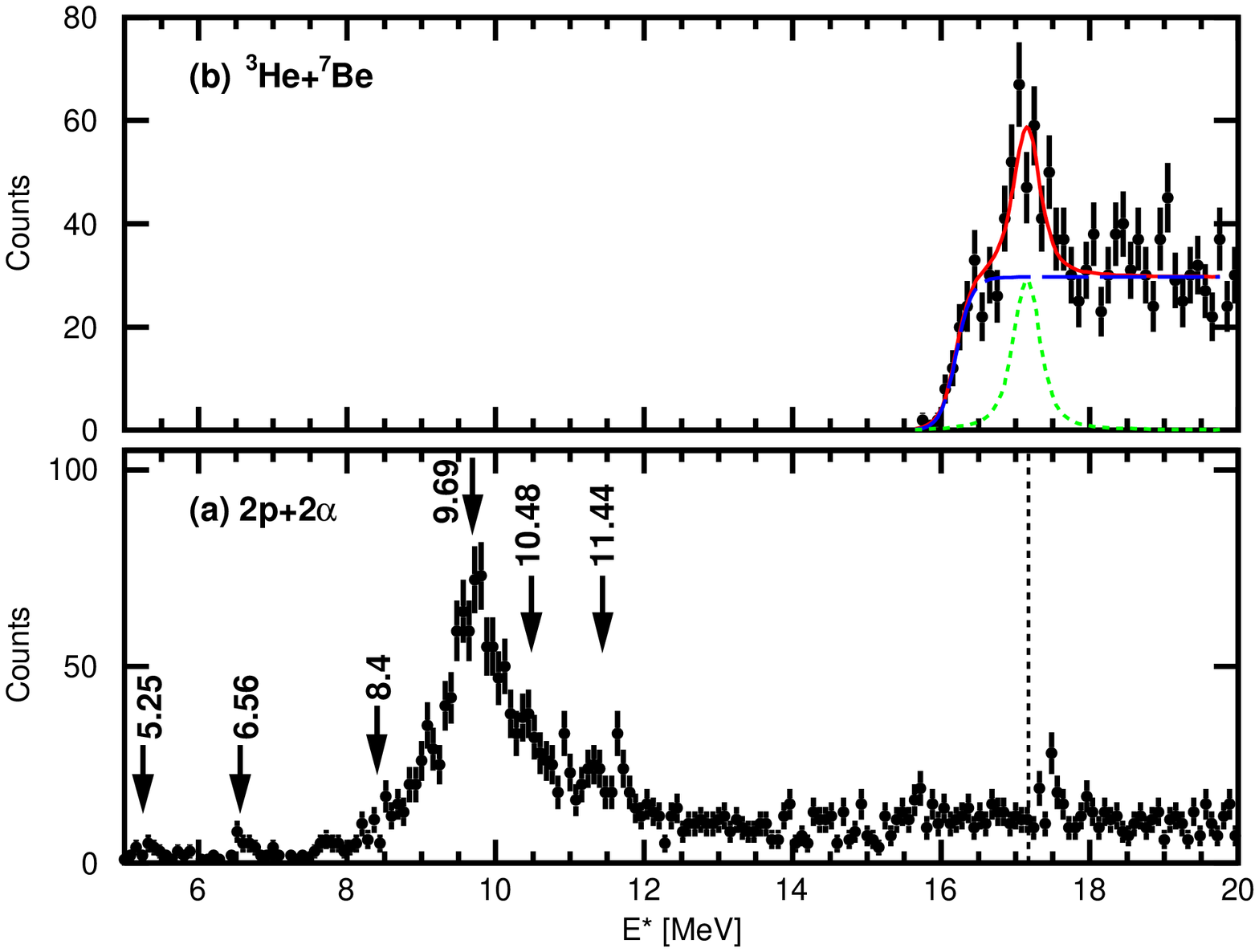}
  \caption{Excitation-energy spectra obtained for  (a) the 2$p$+2$\alpha$ and (b) the $^3$He+$^7$Be exit channels of $^{10}$C. For the four-body exit channel in (a), all detected 2$p$+2$\alpha$ events are included, while in (b), only the transverse decays ( $|\cos\theta|<0.2$) are used in constructing the spectrum. A fit to the $^3$He+$^7$Be data is shown as the solid red line in (b), where the individual Breit-Wigner peaks (modified by the detector resolution) are indicated by the dotted green curve. The dashed blue curve is an estimate of the background. The arrows in (a) show the location of peaks identified in Refs.~\cite{Charity:2008,Charity:2011}.}

\label{fig:C10}
\end{figure*}

The 2$p$+2$\alpha$ and $^3$He+$^7$Be excitation-energy spectra obtained in the neutron pick reactions of this work are displayed in the Fig.~\ref{fig:C10}. The results for the 2$p$+2$\alpha$ channel in Fig.~\ref{fig:C10}(a) are consistent with that obtained at the same bombarding energy and target in Ref.~\cite{Charity:2011} and is dominated by a state at $E^*$=9.69~MeV.  This previous work also identified smaller peaks at 
$E^*$=10.48(20) and 11.44(20)~MeV as indicated by the arrows in Fig.~\ref{fig:C10}(a). These secondary peaks are not so obvious in the present data, but our statistics are lower making them more difficult to discern if present. In addition the location of the 2$p$+2$\alpha$ peaks observed in the $^{10}$C
 inelastic excitation studies are also indicated by the arrows in Fig.~\ref{fig:C10}(a); a doublet at $E^*\sim$5.25~MeV, a triplet $E^*\sim$6.56~MeV, and a broader peak at $E^*$=8.4(1)~MeV. Such peaks are either significantly suppressed or not observed in  this work, consistent with their presumed strong cluster structure. The stronger yield of the 9.69-MeV state indicates it has a more shell-model-like structure.

 In Ref.~\cite{Charity:2011}, the  9.69-MeV state was shown to have $\alpha$+$^6$Be$_{g.s.}$ and  $p$+$^9$B$_{2.34 \textrm{MeV}}$ decay branches in addition to a more unusual 
branch where the $\alpha$-$\alpha$ relative energy is consistent with the $J^{\pi}$=2$^+_1$ $^8$Be resonance, all the $p$-$\alpha$ relative energies are consistent with  $^5$Li$_{g.s.}$ resonances, and the $p$-$p$ relative energy is small reminiscent of a di-proton final-state interaction.  We presume this state is produced from neutron transfer to the $p$-shell and is thus either $J$=0$^+$, 1$^+$, 2$^+$, or 3$^+$. Indeed the emission of a $p$-shell proton should leave the system in a negative-parity state consistent with the significant proton decay branch (17\%) to the $J^{\pi}$=5/2$^-_1$, $E^{*}$=2.34~MeV state of $^9$B \cite{Charity:2011}. 

Based on the known levels in the mirror nucleus $^{10}$Be, the most likely analog is the 9.64-MeV, $J^{\pi}$=2$^+$ state. Note that we are using  the excitation energy from Refs.~\cite{Hamada:1994,Soic:1996,Charity:2008} rather than the compiled value of $E^*$=9.560~MeV \cite{ENSDF}. The width of our $^{10}$C peak ($\Gamma$=490~keV \cite{Charity:2011}) is of similar magnitude but larger than the value of $\Gamma$=141~keV \cite{ENSDF} for the $J$=2$^+$ level in the mirror system which is not unreasonable as the proton-rich member of a mirror pair of levels in the continuum generally has a  larger width.

The $^3$He+$^7$Be excitation energy-energy spectrum for transverse decay, shown in Fig.~\ref{fig:C10}(b), is  dominated by a single 
 peak  at $E^*\sim$17~MeV. This   
peak is associated with decay to the ground state of $^7$Be as no enhancement of the 429-keV $\gamma$ rays associated with the first excited state of $^7$Be was observed in CAESAR. The solid red curve shows a fit to the experimental data with a Beit-Wigner-shaped peak (modified by the detector resolution) and the blue dashed curve is the fitted background contribution. Fitted parameters are listed in Table~\ref{tbl:C10}. The fitted peak energy is $E^*$=17.17(4)~MeV  with 
an intrinsic width consistent with zero [$\Gamma$=57(256)~keV]. There are no known states in the mirror system 
$^{10}$Be close to this energy so no assignment to analog states can be made at present.

\begin{table}
\caption{Fitted mean excitation energies $E^*$, intrinsic widths $\Gamma$, 
and cross sections obtained for the $^{10}$C levels observed in Fig.~\ref{fig:C10}.} 
\label{tbl:C10}
\begin{ruledtabular}
\begin{tabular}{cccc}
$E^*$ & channel& $\Gamma$ & $\sigma_{peak}$ \\
{[}MeV] &       & [keV] & [$\mu$b]  \\
\hline
9.69\footnotemark[1]   &   2$p$+2$\alpha$ & 490\footnotemark[1]  & 369(73) \\    
17.17(4) & $^{3}$He+$^7$Be & 221(117) & 6.9(13) \\

\end{tabular}
\end{ruledtabular}
\footnotetext[1]{from Ref.~\cite{Charity:2011}}
\end{table}

In Fig.~\ref{fig:C10}(a) there is no indication of any decay branch of this state to the 2$p$+2$\alpha$ channel (see dotted line for the energies of the fitted level). However at such large decay energies, the detection efficiency of the 2$p$+2$\alpha$ channel is very small as many of the decay fragments are emitted outside the angular acceptance of the HiRA. The simulated efficiency of detecting all four particles is a factor of 6 smaller than the $^3$He+$^7$Be result with the transverse decay cut ( $|\cos\theta|<0.2$). Combined with a larger simulated experimental resolution (FWHM~700~keV), it is possible that this peak contributes to the observed mostly-flat background at large energies in Fig.~\ref{fig:C10}(a) and thus we cannot rule out that this state  also has a  non-negligible branching ratios  to the 2$p$+2$\alpha$ channel.

\section{Reaction Mechanism}
One might imagine that these transfer reactions are very peripheral and the loosely-bound valence neutron in the $^9$Be target nucleus (separation energy of 1.66~MeV) is 
transferred to the projectile leaving a remnant $^8$Be nucleus is its ground or a low-lying excited state. 
However, the reactions are more complex than that. Information of the remnant target system can be gleaned from reconstructing its excitation energy using energy and momentum conservation from the initial beam momentum and final momenta of the projectile fragments measured in the experiment. By using the term ``excitation energy'' we do not wish to imply that the 8 remnant target nucleons are necessarily left is an excited state of $^8$Be. Rather this term is used to just give the energy of these nucleons in their center-of-mass frame above the $^8$Be ground-state energy. 

\begin{figure}
  \includegraphics*[scale=.4]{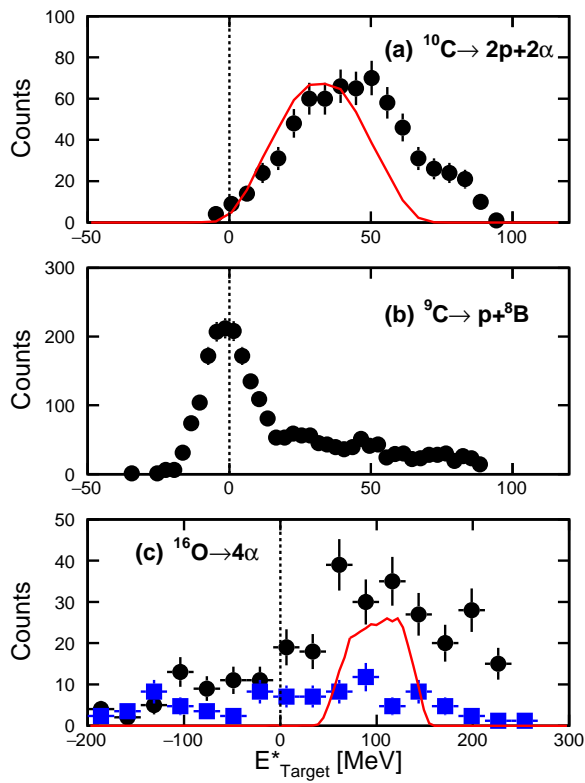}
  \caption{Spectra of the reconstructed excitation-energy of the target nucleus after a transfer or inelastic-scattering reaction. The 
experimental results  are indicated by the data points. The solid-red curves shows the results from the Monte Carlo simulations with an single value of 
 $E^*_{target}$ to indicate the experimental resolution. In (c), the blue-squares show an estimate of the background contribution.}
\label{fig:target}
\end{figure}

The distribution of this energy is plotted in Fig.~\ref{fig:target}(a), as the data points, for the 9.69-MeV state of $^9$C [Fig.~\ref{fig:C10}(a)]. For comparison, the solid curve shows the simulated result (App.~\ref{sec:MC}) for a single value of the  target excitation energy ($E^*_{target}$=33~MeV). The large simulated width is predominantly a result of  the uncertainty in the magnitude of the energy loss of the decay fragments in the target material. We have chosen the 2$p$+2$\alpha$ exit channel for this demonstration, as the energy calibrations of the CsI(Tl) light output are well constrained for these particles and their energy loss in the target is relatively small. Although the simulations explain a significant fraction of the experimental width, the most striking feature is that there is no peak near zero excitation energy and the average is around 40~MeV. This castes doubt on the presumption of the  peripheral nature of these collisions.

 For comparison in Fig.~\ref{fig:target}(b), we show the distribution of $^9$Be target excitation energy associated with inelastic scattering of the $^9$C projectile to its first excited state. The invariant-mass spectrum obtained from the decay of this state to the $p$+$^8$B channel was presented in \cite{Brown:2017}. In this case there is a strong peak at $E^*_{target}\sim$0~MeV and so the inelastic-excitation process has a strong peripheral component that appears to be  lacking for the transfer reaction. We find similar results for the other states formed in the transfer reactions in this work. For example in Fig.~\ref{fig:target}(c), the  excitation-energy distribution for the $^8$Be remnant associated with the 19.262-MeV $^8$Be$_{g.s.}$+$^8$Be$_{g.s.}$ states (Fig.~\ref{fig:fit88}) is shown as the black  circular data points. This peak sits on a significant background and we have used the adjacent low-excitation-energy region to estimated this contribution. The blue-square data points show this contribution after normalizing its magnitude to be consistent with background decomposition in Fig.~\ref{fig:fit88}. This background accounts for most of the yield at  negative values of $E^*_{target}$. However at positive excitation energies,  the distribution above the background is very broad extending up to $\sim$200~MeV. This is 
much broader than the experimental resolution which is indicated by the solid curve which was generated from our simulations with $E^*_{target}$=100~MeV.     

The larger values of $E^*_{target}$ may be a consequence of the large momentum mismatch at the high bombarding energies of this work. For instance this mismatch will be reduced for less peripheral collisions  where the transferred neutron can be placed more in the interior of the projectile. Of course such collisions may also lead to knockout of the projectile's nucleons and other dissipative processes and the events we observed represent a balance between the likelihood of these processes and the difficulty of   momentum matching in peripheral collisions.

In the study of neutron transfer reactions with a $^{22}$Mg fragmentation beam 
using $\gamma$-ray spectroscopy, Gade \textit{et al.} concluded that
the yields obtained with a $^9$Be target were too large to be explained by
the pickup of the weakly-bound $^{9}$Be valance neutron \cite{Gade:2011}. From the measured longitudinal momentum distribution of the final projectile fragments, they also concluded that these transfer reactions were not two body is nature, i.e., the projectile and target after the transfer were not both left in well-defined excited states. In addition they inferred that the reactions with the $^{9}$Be target were dominated by the pickup of one of  the deeply-bound neutrons which would lead to $E^*_{target}>$20~MeV. This is qualitatively consistent with our observations.  Gade \textit{et al.} also studied transfer reaction with a 
$^{12}$C target and found a very different result. Here the yields were found to be consistent with a  two-body reaction mechanism and a coupled-channel-Born-approximation calculation was able to reproduce the measured cross sections.

Finally is it interesting to compare the yields for these transfer reaction to those for other types of reactions we have studied. For our $^{17}$Ne beam, we have also measured neutron knockout \cite{Brown:2014,Brown:2015} and inelastic excitation \cite{Brown:2017}. The  knockout cross sections to the ground and first excited states of $^{16}$Ne are 2.91(9) and 0.92(5)~mb, respectively, both greater than the largest transfer yield of 0.813(18) ~mb for the 5.135-MeV state of $^{18}$Ne. The yield for the 
inelastic excitation of the projectile to its second excited state ($E^*$=1.76~MeV, $J^{\pi}$=5/2$^-$) is even larger at 8.8(2)~mb.  

For the $^{9}$C beam, the largest transfer cross section of 369(73)~$\mu$b is for the 9.69-MeV state in $^{10}$C. In comparison, the cross sections for other simple processes we studied are much larger. The neutron knockout cross section to the ground state of $^8$C \cite{Charity:2011} is 3.8(3)~mb, while the proton knockout cross sections to the first, second, and isobaric analog states of $^8$B \cite{Brown:2014a} are 12.0(20), 42.0(40), and 1.2(1)~mb, respectively. Finally the inelastic scattering cross sections to first, second, and forth excited states of $^9$C \cite{Brown:2017} are 3.74(20), 5.91(40), and 4.12(40) mb, respectively.

The cross sections for these neutron transfer reactions are smaller than other reaction types, even smaller than those for neutron knockout reactions,  which for such proton-rich beams are known to be suppressed  relative to Eikonal-model predictions \cite{Tostevin:2014}. However even in the present studies which were optimized for producing two-proton emitters  via such  knockout reactions, the detected transfer yields were adequate to identify a number of states. Partly this results from the fact that most of these states undergo two-body decay and thus have higher detector efficiencies than the three-body and high-order decays associated with the two-proton emitters.     

\section{CONCLUSION}

We have used invariant-mass spectroscopy with the HiRA and CAESAR arrays to study excited states in the continuum produced in  neutron transfer reactions  to fast secondary beams of $^{9}$C, $^{15}$O, and $^{17}$Ne. With the thick $^9$Be target, which was selected to produced adequate yields with the low beam rates, the  experimental resolution was found to be very sensitive to the orientation of the decay axis of these states. For two-body decays in particular, the best resolution was found for events where the decay axis is perpendicular to the beam direction. Here the uncertainty associated with energy-losses of the decay products in leaving the target material are minimized. 
These transfer reactions were found to leave the remnant target nucleons with  large excitation energies. Futher studies are needed to understand this, but at present this excludes the extraction of spectroscopic factors from comparisons with DWBA calculations.

With the $^{17}$Ne beam, we have confirmed the spin assignments made by Hahn \textit{et al.} \cite {Hahn:1996} for a number of $^{18}$Ne excited states.
 In addition we have found new excited states 
in $^{16}$O and $^{18}$Ne at high excitation energies. Some of these decays are highly fragmented with up to four particles in the continuum. This includes 
an exotic fission mechanism for $^{16}$O states  
resulting in two $^8$Be$_{g.s.}$ fragments. A newly-found  high-$T$ state in $^{18}$Ne was observed to decay to the isobaric analog state in $^{17}$F. The latter was also found to have isospin non-conserving $\alpha$ and proton decay branches. Finally a new excited state in the $^{10}$C was also found.

 This works demonstrates the usefulness of 
invariant-mass spectroscopy in transfer reaction with fast fragmentation beams. Unfortunately, 
cross sections are typically much smaller than other simple reaction mechanisms such as knockout or inelastic excitation.  However, as in the present work, transfer data can be obtained in concert with data from other reactions.

\begin{acknowledgments}
We thank Prof. Alex Brown for lessons on using the OXBASH code.  This material is based upon work supported by the U.S.\ Department of 
Energy, Office of Science,  Office of
Nuclear Physics under Award numbers
 DE-FG02-87ER-40316, DE-FG02-04ER-41320, and DE-SC0014552 and the NSF under grant PHY-156556. K.W.B. was supported by a National Science Foundation Graduate Fellowship under Grant No. DGE-1143954 and J.M. was supported by a Department of Energy National Nuclear Security Administration Steward Science Graduate Fellowship under cooperative Agreement No. DE-NA0002135.
\end{acknowledgments}

\appendix
\section{Monte Carlo Simulations}
\label{sec:MC}
The experimental resolution and detection efficiency were determined from Monte Carlo simulations of the reactions which incorporated the following effects.
\begin{enumerate}
\item  The energy loss of the beam particle and decay fragments in the target material were taken from Ref.~\cite{Ziegler:1985}. The reaction is assumed to occur randomly in depth within the limits of the physical target.
\item Small-angle scattering of the beam particle and decay fragments in the target material following Ref.~\cite{Anne:1988}. 
\item The effect of a realistic beam spot size ($\sim$1~cm diameter) and the known momentum acceptance of the secondary beam are included. 
\item The angle resolution associated with the pixel-size of the Si strip $\Delta E$ detectors are included.
\item The energy resolution of the CsI(Tl) detectors are estimated based on our calibration beams.
\item The detection  efficiency includes the loss due to nuclear reactions of the incident particles with the Cs and I nuclei  in the $E$ detector \cite{Charity:1995,Morfouace:2017}.
 
\item The intrinsic line shapes of resonances were taken to have a Breit-Wigner form with the centroid and width adjusted in the fits unless otherwise specified. 
\end{enumerate}
The Monte Carlo events produced by the simulation are analyzed in the same manner as the experimental data. The ingredients in the simulations were fine tuned by fitting known narrow resonances. For example, the $p$+$^{17}$F resolution was fine tuned by fitting the  2$p$+$^{15}$O resonance peak associated with the decay of the second excited state of $^{17}$Ne as discussed in the \cite{Brown:2014,Brown:2015}. Both transverse and longitudinal decays are considered as these have sensitivities to different ingredients. For the fission of $^{16}$O states into two $^8$Be$_{g.s.}$ fragments producing a final exit channel of four $\alpha$ particles, three resonances were used for fine tuning. These are the $^{8}$Be$_{g.s.}\rightarrow$ 2$\alpha$ resonance plus the 3$\alpha$ resonances associated with the $^{12}$C second (Hoyle state) and third ($J^{\pi}$=3$^-$) excited states.  

Input primary angular and velocity distributions of the parent fragments formed in the transfer reactions  were adjusted so that reconstructed secondary distributions (obtained from the decay fragments after the effects of the detector acceptance and resolution are incorporated) match their experimental counterparts. For asymmetric exit channels like $p$+$^{17}$F, these is an uncertainty in extrapolating to zero degree as the detection efficiency vanishes here and this adds uncertainty to our final  cross sections. However, as the $d\sigma/d\theta$ must vanish as one approaches zero degrees, this uncertainty is not large. We estimate this uncertainty is less than 15\%. For the $^{16}$O fission channels, this zero degree region is sampled by the experimental events so a similar problem does not exist.

%

\end{document}